\documentclass[3p, preprint]{elsarticle}
\usepackage{microtype, xcolor}
\usepackage{graphicx}
\usepackage{subfig}
\usepackage{booktabs} 

\usepackage{ulem}

\usepackage{hyperref, nicefrac,bm}

\usepackage{amsmath}
\usepackage{amssymb}
\usepackage{mathtools}
\usepackage{amsthm, dsfont, enumerate}
\usepackage{wrapfig, caption}

\newtheorem{theorem}{Theorem}
\newtheorem{corollary}[theorem]{Corollary}
\newtheorem{definition}{Definition}
\newtheorem{example}{Example}
\newtheorem{remark}{Remark}
\newtheorem{assumption}{Assumption}
\newtheorem{lemma}{Lemma}

\newcommand{\act}{{\bf a}}
\newcommand{\underdelta}{\underline{\delta}}
\newcommand{\tgam}{ \tilde{ \gamma}}
\newcommand{\ug}{{\underline g}}
\newcommand{\og}{{\bar g}}
\newcommand{\bbP}{{\mathrm P}}

\newcommand{\ospgame}{unified seller game}

\newcommand{\X}{{\mathcal X}}
\newcommand{\Z}{{\mathcal Z}}
\newcommand{\N}{{\mathcal N}}
\newcommand{\F}{{\mathcal F}}
\renewcommand{\l}{{\bar l}}
\newcommand{\barp}{{\bar p}}
\newcommand{\bara}{{\bar a}}
\renewcommand{\P}{{\mathcal P}}
\newcommand{\U}{{U}}
\newcommand{\B}{{B}}
\renewcommand{\b}{{b}}
\newcommand{\A}{{\mathcal A}}
\renewcommand{\L}{{\mathcal L}}
\newcommand{\lx}{{\bf{x}}}
\newcommand{\fstar}{f^*}
\newcommand{\sstar}{{*}}
\newcommand{\zsstar}{\z^\sstar}
\newcommand{\hlx}{\widehat{{\bf x}}} 
\newcommand{\hlxe}{\widehat{x}} 
\renewcommand{\a}{{\bf a}}
\newcommand{\Arg}{\mbox{Arg}}
\DeclareMathOperator*{\Argmax}{Arg\,max}
\DeclareMathOperator*{\Argmin}{Arg\,min}
\DeclareMathOperator*{\argmax}{arg\,max}
\DeclareMathOperator*{\argmin}{arg\,min}

\newcommand{\uu}{\underline{u}}
\newcommand{\ou}{\bar{u}}
\newcommand{\E}{\mathcal{E}}
\newcommand{\be}{{\bar \eta}}
\newcommand{\teta}{{\tilde \eta}}
\newcommand{\prob}{\mathbb{P}}
\newcommand{\I}{\mathcal{I}}
\newcommand{\D}{\mathcal{D}}
\renewcommand{\Z}{\mathcal{Z}}
\newcommand{\p}{{\bf p}}
\newcommand{\z}{{\bf z}}
\newcommand{\psstar}{\p^\sstar}
\newcommand{\bdelta}{{\bm \delta}}

\newcommand{\dirac}[1]{ {\Delta({ #1})}}
\newcommand{\red}[1]{ {\color{red}{ #1}}}
\newcommand{\tus}[1]{{\color{blue}{#1}}}
\newcommand{\rev}[1]{{\color{black}{#1}}}

\newcommand{\indic}[1]{\mathds{1}_{\left\{ #1 \right \}} }
\newcommand{\ignore}[1] {}
\newcommand{\extraproof}[1] {}

\begin{document}

\begin{frontmatter}

\title{Strategic Pricing and Ranking in Recommendation Systems with Seller Competition}
\author[aut_label_tus]{Tushar Shankar Walunj} \ead{tusharwalunj@iitb.ac.in}
\author[aut_label_kav]{Veeraruna Kavitha} \ead{vkavitha@iitb.ac.in}
\author[aut_label_jkn]{Jayakrishnan Nair} \ead{jayakrishnan.nair@ee.iitb.ac.in}
\author[aut_label_shik]{Priyank Agarwal} \ead{22n0453@iitb.ac.in}
\affiliation[aut_label_tus, aut_label_shik, aut_label_kav]{organization={Industrial Engineering and Operations Research, IIT Bombay},
            addressline={Powai}, 
            city={Mumbai},
            postcode={400076}, 
            state={Maharashtra},
            country={India}}
\affiliation[aut_label_jkn]{organization={Electrical Engineering, IIT Bombay},
            addressline={Powai}, 
            city={Mumbai},
            postcode={400076}, 
            state={Maharashtra},
            country={India}}     
\begin{abstract}

We study a recommendation system where sellers compete for visibility by strategically offering commissions to a platform that optimally curates a ranked menu of items and their respective prices for each customer. Customers interact sequentially with the menu following a cascade click model, and their purchase decisions are influenced by price sensitivity and positions of various items in the menu. We model the seller-platform interaction as a Stackelberg game with sellers as leaders and consider two different games depending on whether the prices are set by the platform or prefixed by the sellers. 

It is complicated to find the optimal policy of the platform in complete generality; hence, we solve the problem in an important asymptotic regime. In fact, both the games coincide in this regime, obtained by decreasing the customer exploration rates $\gamma$ to zero (in this regime, the customers explore fewer items). Through simulations, we illustrate that the limit game well approximates the original game(s) even for exploration probabilities as high as 0.4 (the differences are around 2.54$\%$). Further, the second game (where the sellers prefix the prices) coincides with the approximate game for all values of $\gamma$. 

The core contribution of this paper lies in characterizing the equilibrium structure of the limit game. We show that when sellers are of different strengths, the standard Nash equilibrium does not exist due to discontinuities in utilities. We instead establish the existence of a novel equilibrium solution,  namely `$\mu$-connected equilibrium cycle' ($\mu$-EC), which captures oscillatory strategic responses at the equilibrium. Unlike the (pure) Nash equilibrium, which defines a fixed point of mutual best responses, this is a set-valued solution concept of connected components. This novel equilibrium concept identifies a Cartesian product set of connected action profiles in the continuous action space that satisfies four important properties: stability against external deviations, no external chains, instability against internal deviations, and minimality. 
We extend a recently introduced solution concept \textit{equilibrium cycle} to include stability against measure-zero violations and avoid some topological difficulties to propose $\mu$-EC.
\end{abstract}

\end{frontmatter}

{

\section{Introduction}

Platforms like Amazon and Walmart, which facilitate interaction between buyers and sellers, increasingly rely on curated recommendation systems to enhance the user experience and maximize revenue. In such systems, \textit{`seller visibility'}, often implemented through a ranked menu of recommended items, plays a crucial role in shaping customers' behavior and purchase decisions regarding any seller's product. Naturally, sellers compete for this visibility by adopting strategic behaviors that can influence their ranking.

While classical models focus on optimizing rankings or prices in isolation (e.g., \cite{kempe2008cascade, dhingra2021multi}), modern digital marketplaces face a more intricate challenge: how to jointly align \textit{pricing, ranking, seller incentives, and customer behaviour}, especially in settings where the sellers actively respond to and attempt to influence the platform policies. This interaction leads to a two-sided strategic tension---between platforms optimizing revenue and sellers seeking favorable positioning, which remains under-explored in the recommendation systems literature.

The paper investigates this tension by modelling the interaction between sellers and a revenue-maximizing platform. We focus on a commission-based design, where sellers offer monetary incentives (commissions) to the platform in the hope of securing higher visibility. The platform, in response, constructs a personalized ranked menu and depending on the setting, may also determine  the item prices. It optimizes its revenue by accounting for both the offered commissions and customer click-through behaviour (e.g., as in \cite{kempe2008cascade,dhingra2021multi}).

This interplay is modelled using a Stackelberg game framework with sellers as leaders competing for customer attention by strategically choosing commissions for the platform. We consider two distinct  scenarios:
\begin{enumerate}
\item {Order and Pricing (OP) game:} Sellers (leaders) choose commissions, and the platform (follower) jointly sets the ranking and item prices.
\item {Order at Set-Prices (OS) game:} Sellers first fix item prices and then choose commissions. The platform, observing the commissions, determines only the ranking.
\end{enumerate}
These models capture a wide spectrum of real-world platforms—from those retaining full control over prices to marketplaces where prices are fixed externally (e.g., MSRP or multi-platform listings). 
\rev{In the current platform-aided marketplaces, the sellers set the prices, however the prices are significantly influenced by the norms and the policies of platform (e.g.,  see \cite{Walmart, Flipkart} on how major platforms like Walmart, Flipkart influence seller prices). This motivates us to consider the above two games and analyse them individually as well as to obtain a comparative study.}
Further, unlike the existing literature that often decouples pricing and ranking decisions, our model integrates these choices. We thus uncover a rich feedback loop: seller commission strategies influence menu placement, which in turn affects the customer choices and the platform revenue, thereby feeding back into the seller's revenue and hence the strategic considerations. This strategic interdependence gives rise to novel equilibrium behaviour not captured by classical solution concepts.
Our work makes the following key contributions:
 \begin{itemize}
     \item \textit{Stackelberg game formulation}:
    We formalize the interaction between strategic sellers and a revenue-maximizing platform as two types of Stackelberg games, corresponding to different realistic scenarios in platform markets. 
    
\item  \textit{Computational framework for platform optimization}: We pose the platform’s response as an MDP (Markov Decision Process) and solve it via backward induction. When the customer drop-off rate is high, we derive the closed-form solutions for platform optimization problem, thereby reducing the complexity from exponential to linear in menu size,  and thus making our model scalable.

\item \textit{Unified analysis in low-$\gamma$ regime}: When customers are less likely to explore items deeper in the menu, the OP and OS games converge to  a  common game. This regime enables a unified analysis of seller competition. We term this combined model the \ospgame.

\item  For the scenarios with equal sellers (in terms of reputation and production cost), we establish the existence of pure-strategy NE, where the sellers operate at marginal costs (similar to standard Bertrand duopoly game).

\item \textit{Equilibrium characterization via $\mu$-connected equilibrium cycle}:  
    Our central theoretical innovation is the introduction of a novel set-valued solution concept: the $\mu$-connected Equilibrium Cycle ($\mu$-EC). In games with discontinuities, caused here by rank jumps in menus, the classical pure Nash equilibria does not exist under certain scenarios. The $\mu$-EC captures stable yet cyclic equilibrium for such case-studies: sellers perpetually adjust their commissions in reaction to competitors, but the strategy profiles remain confined within a minimal, recurrent set. We establish  the existence of $\mu$-ECs as a solution 
    for the scenarios  with asymmetric sellers.  Interestingly, the weaker sellers act as a threat that constrains the top-ranked sellers to offer commissions higher than the weak sellers' potential.
    
    This kind of cyclical strategic behaviour mirrors phenomena observed empirically in auction and advertising systems (e.g., \cite{zhang2011cyclical}), where sellers repeatedly adjust bids without convergence.  Our $\mu$-EC formalism captures and generalizes this behaviour in the context of recommendation systems with strategic sellers.

\item 
\rev{The new notion $\mu$-EC is a generalization of the recently proposed solution concept, namely Equilibrium cycle (\cite{EC_Paper}): the  $\mu$-EC  allows 
measure-zero exceptions and avoids some topological difficulties.
It identifies a connected set of action profiles that are (i) stable against external deviations, (ii) contain no external chains, (iii) unstable against internal deviations, and (iv) are minimal. The key refinement lies in permitting temporary deviations that leave the set but quickly return, ensuring that long-run dynamics remain confined within the $\mu$-EC.}

 \end{itemize}


\subsection*{Literature review}
We adopt the cascade click model introduced by \cite{kempe2008cascade} and extended in \rev{\cite{dhingra2021multi, joint_RS_2022,derakhshan2022product, najafi2024multiproduct}} to capture customer interactions with ranked menus. These models, while foundational for customer behaviour modelling, do not incorporate platform-driven pricing or strategic seller behaviour. In contrast, our work embeds these behavioural dynamics within a game-theoretic framework involving commission competition and platform optimization. 
Broadly, recommendation systems have been studied through algorithmic, behavioral, and learning-theoretic lenses, often focusing on optimizing rankings, prices, or assortments based on buyer preferences. However, such models typically treat sellers as passive entities and overlook their potential to strategically influence platform decisions. Our work departs from this by treating sellers as active agents who compete via commissions, directly affecting both ranking and pricing outcomes.

\rev{On the other hand, authors in (\citep{ks2023pricing, saberi2019stackelberg, chen2021assortment}) have studied platform-seller strategic interactions using Stackelberg game formulations, where sellers are the suppliers and the platform is a retailer. These strands of literature do not consider the design of menus that strategically present the products of multiple sellers to customers in an optimal order. By contrast, we consider both pricing and ranking decisions in two distinct (OS and OP) games.

Thus, we have pieces of work that focus only on ranking decisions (or optimal menu design) and other strands that consider pricing games alone. But to the best of our knowledge, we did not come across any literature that captures the framework of joint decisions, which is predominant in real-world applications. Our work, therefore, fills this gap. At the same time, it is loosely related to several other themes of work reported in the literature, a few of which we discuss below.

In the recent past, there has been a significant body of work that discusses the aggressive dynamic prices set by sellers using automated algorithms. The use of algorithms in place of human decision-making increases price fluctuations, as demonstrated in \cite{chen2016empirical, johnson2023platform, amazon_automate_pricing, hanspach2024algorithms}.  For example, \cite{chen2016empirical} states that even in 2015 about one-third of Amazon’s top-selling products were already priced using algorithms, with algorithms changing prices hundreds of times more often than human sellers. Authors in \cite{johnson2023platform} show that platforms can influence these outcomes through their ranking rules and by rewarding price cuts. Overall, these studies show that platforms are not just neutral intermediaries, but they actively shape the competition through their decisions. This line of work confirms that sellers also behave strategically, providing strong motivation for our framework which explicitly incorporates strategic sellers into recommendation systems.

Another related branch of literature \cite{qiu2024can, chen2024platform} discusses platforms that directly compete with sellers on certain products, raising fairness concerns. While the platform in our model does not directly compete with sellers for market share, our framework can easily be extended to study such scenarios by setting commission for in-house products to 100 \%, and allowing the remaining sellers to  compete as before. 

The role of commissions in platform-based markets has also been studied extensively in the literature, for example, \cite{chen2019commission, hasiloglu2021analysis, li2025regulating}. Authors in \cite{hasiloglu2021analysis} analyze a cooperative setting with two sellers and a platform, where pricing, service quality, and commissions are jointly determined. In their framework, the platform benefits from intense seller competition, much like in our model, while the sellers prefer cooperation to increase their profits.
However, these pieces of work do not consider the influence of the ranking  decisions.
Further, unlike such cooperative scenarios, our framework is fully non-cooperative: sellers choose commissions purely to maximize their own revenue, and the platform optimizes its policy in response. This non-cooperative setting introduces new challenges. In particular, we encounter situations where a classic pure-strategy Nash equilibrium does not exist due to discontinuities in seller’s payoff.
As already mentioned, to address this, we introduce the concept of $\mu$-EC,  that can capture cyclic strategic outcomes.

This work also refines the recently introduced cyclic equilibrium outcome, namely Equilibrium cycle introduced in \cite{EC_Paper}, by proposing a refined $\mu$-EC. In this regard, we now list below strategic cyclic phenomenon observed in other applications. 
In related empirical contexts such as airline and gasoline markets, researchers have documented recurring price wars known as \textit{Edgeworth cycles}. These cycles describe a pattern where firms repeatedly undercut each other until prices fall close to marginal cost, after which prices reset to higher levels and the cycle restarts \cite{edgeworth_cycle_airlines, noel2007edgeworth, noel2011edgeworth}. Similar dynamics also have been observed in online advertising, where the competing bidders adjust bids in persistent cycles without convergence \cite{zhang2011cyclical}. The persistence of such dynamics highlights the limitations of classical equilibrium notion of Nash equilibrium in explaining real-world competitive outcomes. Our $\mu$-EC framework offers a natural theoretical explanation for these phenomena by formally capturing oscillatory yet stable strategy profiles.

In summary, our work brings together insights from cascade models of consumer behavior, pricing and ranking optimization, seller–platform competition, algorithmic pricing, and equilibrium analysis in discontinuous games. 

}

This paper is structured as follows: Section~\ref{sec:Problem_description} presents the system model. Section~\ref{sec_MDP} discusses the platform’s optimization   via an MDP framework. We discuss  small-$\gamma$ approximation in Sections~\ref{sec_small_gamma_game} and~\ref{sec_small_OS}. Section~\ref{sec:equilibrium} defines the solution concept $\mu$-EC. 
We conclude in Section~\ref{sec:conclusions}.

\section{Problem description} \label{sec:Problem_description}

We consider a system where the market is mediated by a double-sided platform that facilitates efficient matching between multiple sellers and potential buyers. Sellers use the platform to reach a broader customer base and offer their products, while the customers visit the platform to explore a wider range of products before purchase. For each successful transaction, the seller pays a commission to the platform as compensation for facilitating the match. The platform, in turn, recommends a curated subset of sellers (a menu) based on customer preferences, particularly their price sensitivity, thereby enhancing the shopping experience. Customers interact with the displayed menu and decide whether to purchase an item, upon which the platform ensures the transaction is completed smoothly. Since visibility on the menu designed by the platform directly affects the sales, the sellers compete by strategically setting commissions to improve their chances of being selected  and securing a better position in the menu.

We begin by formally describing the model, its components, and the associated notation.

\subsection{Platform-seller interaction}

Let $\I = \{1, 2, \cdots, N\}$ denote the set of sellers on the platform. Each seller $a \in \I$ offers a product and commits to a `commission' $\delta_a \in [0, 1]$, representing the fraction of the sale price paid to the platform upon a successful transaction. The platform observes these commissions and constructs a `customized menu' (ranked list of $M$ items) for each customer by selecting a subset of $M \leq N$ sellers. This decision also considers the customers' responses to various products or sellers.
This customized menu is represented as a permutation $\pi = (i_1, i_2, \cdots, i_M)$ of sellers, along with corresponding prices $\p = (p_1, p_2, \cdots, p_M)$ assigned to their products.
The customers are assumed to interact with the menu sequentially, from top to bottom. At each step, they either decide to purchase the item displayed at that position, leave the system, or explore the next item in the menu. The platform's expected revenue depends on the menu ordering $\pi$, the price vector $\p$, the commission profile $\bdelta$, and $\gamma$, the probability with which a typical customer decides to explore the next item.

The revenue of any seller depends on its position in the menu (and, of course, on the price set for its item), which in turn depends on: (i) the price of its item,
(ii) the commission it sets,
(iii) its reputation or the product's price sensitivity, and
(iv) the probability that a customer reaches and purchases the item after skipping earlier items. The last aspect directly depends on the position it occupies in the menu designed by the platform, and that design is influenced by the first three aspects.

\subsection{Customer behaviour}

Customers follow a \textit{cascade click model} \citep{kempe2008cascade} while interacting with the menu. Starting from the top, any typical  customer is assumed to  evaluate each item and make one of the following decisions for each of them:
\begin{enumerate}
    \item \textit{Purchase} the product from seller $a$ at the offered price $p_a$ with probability $\beta_{a}(p_a)$,
    \item \textit{Skip and continue} to the next item with probability $(1 - \beta_{a}(p_a))\gamma$, or 
    \item \textit{Exit} the system entirely with  probability $(1 - \beta_{a}(p_a))(1 - \gamma)$. 
\end{enumerate}
If a customer purchases from seller $a$, the platform earns revenue $\delta_{a} p_a$, while the seller earns $(1 - \delta_a) p_a$, and the process terminates. Therefore, the expected revenue for the platform is determined by the cumulative probability of purchases across the menu, weighted by the commissions and the prices set. For example, the revenue contribution from the $k$-th position in $\pi$ depends on the probability that the customer reaches position~$k$ and makes a purchase, which is recursively influenced by the skipping and non-exiting probabilities of the preceding items.

The attraction probability function $\beta_a(\cdot)$ captures the price sensitivity of customers toward the product of seller~$a$. We make the following assumption\footnote{This assumption that the customer response function—specifically, the probability of a price being acceptable—is differentiable and decreasing is standard in the literature (e.g., see~\cite{rader1973nice_demand_function}, \cite{huang2013demand_all}, \cite{demand_f_1}). As the price increases, the probability of acceptance decreases significantly, indicating that customers become more price-sensitive.} on the function $\beta_a(\cdot)$, which is commonly considered in the literature in similar contexts (e.g., \cite{joint_RS_2022},~\cite{walunj2023interplay}).

\begin{assumption}\label{assumption_beta_fun}
The function $\beta_a(\cdot)$ is differentiable and strictly decreasing. Moreover, for all $p \in [0, p^{\text{max}}]$, we have $0 < \beta_a(p) \leq 1$ and $\beta_a(0) = 1$, where $p^{\text{max}}$ is a sufficiently large upper bound on feasible prices. Additionally, for each $a \in \I$ and for any $c \ge 0$, there exists unique optimizer $p^*(a) \in \Argmax_p\beta_a(p) (p-c)$. 
\end{assumption}
\textit{In this paper, `$\Argmax$' denotes the set of maximizers, while `$\argmax$' refers to the unique maximizer, once uniqueness is established.}

\subsection {Decisions of various agents}
The agents make three key decisions in this model:  commission selection, menu design, and price setting. The sellers strategically choose the commission for the platform in each successful (matching) transaction. Next, the platform chooses the menu, or permutation $\pi$, to maximize its expected revenue. It is important to note that a product with a higher probability of being purchased may not always be prioritized by the platform---the platform's gain from such a transaction could be small due to a lower commission value.
Finally, the prices of the items must be set appropriately to maximize the revenue of the involved agents.
We consider two models for setting prices: either the platform sets the prices for all items, or the sellers set the prices for their own items.

This aspect of the problem has a close resemblance to the well-known Bertrand duopoly, particularly when sellers set the prices of their items. However, the optimization and game-theoretic challenges considered in this paper are significantly more complex, as they are interlaced with the menu design problem. This problem modulates customer responses in a different manner from the standard duopoly or oligopoly setting. Its complexity is significantly more, even compared to the stand-alone menu selection problem, which itself is regarded as a difficult problem in the domain of discrete optimization (see \citep{ceberio2012review, dhingra2021multi}).

In the marketspace, one can typically observe scenarios where the sellers pre-set the prices, and the platform only designs the menu \cite{chen2016empirical, dirusso2011determinants, chen2021reputation}; while, in some other cases, the platforms may set both the prices and the menu-order (platforms like Amazon price as well as ranks their own products \citep{amazon_wikipedia, farronato2023self}). As already mentioned, we study both the scenarios and accordingly have two different Stackelberg games.

\begin{enumerate}
    \item 
    In the first scenario, termed the Order and Pricing (OP) game, the sellers act as leaders, strategically setting their commissions $\bdelta$ to maximize individual profits. The platform, as the follower, then selects the optimal ranking $\pi$ and price vector $\p$ based on the offered commissions $\bdelta$.
    The platform’s decisions depend on the commissions, the sellers’ reputations (as reflected in the price sensitivity function), the trade-off between higher profit shares (driven by commissions and prices), and the potential purchase probabilities.
    \item In the second scenario, the Order at Set-Prices (OS) game, the sellers pre-fix the prices of their individual products to maximize their per-sale revenue. Next, acting as leaders, they set commissions $\bdelta$, anticipating the platform's ranking decision; their decisions can also depend upon the pre-fixed prices. Finally, the platform (follower) selects the optimal permutation $\pi$ based on the prices set by the sellers and the commissions offered.
\end{enumerate}

\subsection{Utilities of different agents}
Before defining the different games in our model, we first specify the utilities of the agents for the given decision variables $(\bdelta, \pi, \p)$. 
In the following sections, we will observe that the platform selects a subset $\lx \subseteq \I$ of sellers, with $|\lx| \ge M$, to be included in the menus shown to different customers (this is considered for fairness purposes). It then chooses an ordered subset of length $M$ from the sellers in $\lx$ for each customer in a random (and fair) manner. This ordered subset precisely forms the menu $\pi$ shown to that customer. We will discuss random menus in more detail later; for now, we focus on the utilities under a fixed menu $\pi$.

For given $(\bdelta, \pi, \p)$, the revenue function of the platform (follower) is:

\begin{equation} \label{eqn_revenue_function}
R(\bdelta, \pi, \p) = \sum_{j = 1}^M p_{\pi(j)} \delta_{\pi(j)}   \underbrace{\left( \beta_{\pi(j)}(p_{\pi(j)}) \prod_{k < j} \left(1 - \beta_{\pi(k)}(p_{\pi(k)})\right)\gamma \right)}_{\text{Probability of reaching and purchasing at position } j} .    
\end{equation}
Here, the term $\prod_{k < j} \left(1 - \beta_{\pi(k)}(p_{\pi(k)})\right)\gamma$ represents the probability of a customer reaching the $j$-th position item in the permutation $\pi$, after skipping all previous items and not exiting.
For the platform, the higher shares~$\delta_a p_a$ may increase the revenue per sale, but they may also reduce the purchase probabilities~$\beta_{a}(p_a)$, while the ranking~$\pi$ influences the visibility of items of different sellers, for which the sellers compete.

For example, if $\lx = \{1,2\}$, $\bdelta = (\delta_a, \delta_b)$, $\pi = (a,b)$ and $\p = (p_a, p_b)$ then the revenue of the platform is: 
$$
R(\bdelta, \pi, \p) = p_a \delta_a \beta_a(p_a) + \gamma \left (1-\beta_a(p_a) \right )  p_b \delta_b \beta_b (p_b ), 
$$ 
while that of the sellers are $(1-\delta_a) p_a  \beta_a(p_a) $ and $(1-\delta_b) p_b  (1-\beta_a(p_a)) \gamma  \beta_b (p_b )$ respectively. These seller revenues do not include the respective production costs. We now describe the utility functions of the sellers, including the costs.
The profit of seller $a$ upon selling its product is given by $(1 - \delta_a) p_a - c_{m_a}$, where $c_{m_a}$ is the production cost. However, the opportunity to sell its product depends on $(\pi, \p)$, and the combination of these two factors defines its overall utility. 
\rev{For given $(\bdelta, \pi, \p)$,  the utility of seller $\rev{a \in \I}$ 
equals:
\begin{eqnarray}\label{Eqn_seller_game_general_utility}
\U_a (\bdelta, \pi, \p) =  
    ( (1 - \delta_{a}) p^*_{\pi,a}  - c_{m_a} ) \beta_{a}( p^*_{\pi,a}) \prod_{j < k} \left(1 - \beta_{\pi(j)}(p^*_{\pi, \pi(j)} \right)\gamma,  
    \quad \mbox{ if } a = \pi(k)  \mbox{ for some } k \le  M,
\end{eqnarray}%
i.e., if placed in $k$-th position with $k \le M$, 
otherwise $\U_a (\bdelta, \pi, \p) = 0$.}
Next, we formally define the two games.

\subsection{Game formulations}
We begin by defining the OP game.

\subsubsection{Order and price (OP) game} \label{subsec:OP_Game}
In this game, the sellers take the lead by strategically setting commissions  $\bdelta:=\{\delta_a\}_a$ to maximize their own profits. The platform responds by choosing the ranking (or menu) $\pi$ and setting the prices $\p$ for each seller's item, based on $\bdelta$. 
The
sellers face a trade-off: a higher $\delta_a$ incentivizes the platform to prioritize their product (e.g., through a better ranking), but it also reduces their per-sale profit as seen from \eqref{Eqn_seller_game_general_utility}.

Formally,   this scenario can be captured by a Stackelberg game  with two stages:  
\begin{enumerate}
    \item Sellers’ Stage (Leaders): Sellers simultaneously choose commissions,   $\bdelta$.  
    \item Platform’s Stage (Follower): The platform observes $\bdelta$ and selects $\z :=  (\pi, \p)$ optimally to maximize $R(\z; \bdelta)$.  
\end{enumerate}
We begin with detailing the optimization problem  ${\cal P}_F (\bdelta)$ of the follower (the platform) for any given strategy profile $\bdelta$ of the leaders or the sellers (see \eqref{eqn_revenue_function}). Informally, this optimization problem can be posed as:
\begin{equation}\label{eqn_optimization_prob_RS}
\begin{aligned}
\mbox{{\bf   ${\cal P}_F  (\bdelta)$: }}
 \max_{\z=(\pi, \ \p)} R(\z; \bdelta),   
 \text{ subject to } \mbox{ ordering policy}~\pi, \mbox{ price vector } \p \in [0, p^{\text{max}}]^M.
\end{aligned}    
\end{equation}
It is important to note that solving this joint optimization problem for large $N$ is challenging due to its mixed combinatorial and continuous nature. The number of possible permutations of length $M$ is $\binom{N}{M} M!$, and for each permutation, the platform must determine the optimal pricing strategy. Instead, the problem can be posed as a sequential decision problem. More precisely, it can be framed as a Markov Decision Process (MDP), which allows us to derive the optimal policy for the platform (see details in Section~\ref{sec_MDP}). Using this approach, the time complexity can be reduced to approximately $\binom{N}{M} 2^M$, representing a significant reduction. This typecasting offers many advantages:
\begin{itemize}
    \item [(i)] it simplifies the solution, because of the well-known dynamic programming (DP) equations that solve the MDPs (more details in Section~\ref{sec_MDP}); 
    
    \item[(ii)]   the offline methods of MDP can be used to obtain a possibly randomized optimal policy, which further incorporates `fair' allocation of menu-positions to different sellers; the implementation of such a policy would imply display of random IID (identical and independently distributed) menus to each of the customers, details of the same are in Subsection \ref{subsec:implementation_of_policy}; 

    \item [(iii)] finally, such a formulation can organically lead to efficient online learning algorithms, when some of the system parameters are not known.
\end{itemize}

The problem in \eqref{eqn_optimization_prob_RS} for any given $\bdelta$, when posed as an MDP, has a solution $\z^*(\bdelta)$, which is derived in \eqref{eqn_DP_equations}-\eqref{Eqn_opt_policy} of the next section. In some cases, multiple optimizers may exist, but we select an unique `fair' version, where the policy at any stage chooses uniformly among all sellers with `equal utility' (see \eqref{Eqn_opt_policy}). Furthermore, for each~$\bdelta$, the optimal (and fair) policy can be represented as a random menu and price vector, $\z^*(\bdelta) = (\Pi^*(\bdelta), P^*(\bdelta))$, as described below: 
\begin{definition}{\bf[Random menu and price-vectors $(\Pi^* (\bdelta), P^*(\bdelta))$]}
\label{defn_policy} is the policy:
\begin{itemize}
       \item where $\Pi^*(\bdelta)(\pi)$ represents the probability of choosing permutation/menu $\pi$, and
    \item  $\bbP^*(\bdelta)  =  (p^*_{\pi, a} )_{\pi, a}$ represents the prices-vectors, where $p^*_{\pi,a}$ is the price at which the item of seller $a$ is offered, when its item is placed according to menu $\pi$.  
\end{itemize}
    
\end{definition}

This policy $\z^* (\bdelta)$ is actually derived in Section~\ref{sec_MDP}; for now, we assume its existence and proceed towards providing the complete description of the Stackelberg (OP) game.
At the top, we have a strategic form game between $N$ sellers. Each seller chooses a commission $\delta_a \in [0,1]$, and  the response of the follower platform,  $\z^*(\bdelta)$, influences their utilities   (see \eqref{Eqn_seller_game_general_utility}, $\indic{ a =  \pi(k)} $ is indicator of  the seller being placed in $k$-th position in $\pi$): 
\begin{eqnarray}
\label{Eqn_OP_game}
\U_a (\bdelta):=    \sum_{\pi}  \Pi^*(\bdelta) (\pi)  \sum_{k=1}^M \left (  \indic{ a =  \pi(k)} 
( (1 - \delta_{a}) p^*_{\pi,a}  - c_{m_a} ) \beta_{a}( p^*_{\pi,a}) \prod_{j < k} \left(1 - \beta_{\pi(j)}(p^*_{\pi, \pi(j)} \right)\gamma  \right ) \forall \ \rev{ a \in \I}.
\end{eqnarray}

\subsubsection{OS game: order at set-prices}  
In the Order at Set-Prices (OS) game, the sellers decide the prices of their items $\p$ prior to the game; basically, they approach the platform with the items to be sold and with set prices.  
The strategic interaction between the sellers and the platform is once again initiated by the sellers by choosing commissions $\bdelta$; the platform responds now by choosing  (only) an optimal arrangement of the sellers' items or the menu $\pi$. This is applicable to scenarios where the sellers deal with multiple platforms and probably also with multiple retailers. 

The sellers pre-compute their optimal prices based solely on the market response to their own item and set a common price for their item (common across various retailers/platforms). 
Unlike the OP game, where the platform optimizes prices, here, each seller $a$ first determines the price $p^*_a$  (unique by assumption \ref{assumption_beta_fun}) that maximizes their stand-alone revenue, by solving: 
\begin{equation}\label{eqn_item_prices_OS_game}
p^*_a = \argmax_{p \in [0, p^{\text{max}}]} \beta_a(p)  p.  
\end{equation}
This price $p^*_a$ is fixed and remains unchanged during competition. 
Sellers then strategically set $\delta_a \in [0,1]$ to maximize their utility, anticipating the platform’s menu optimization.

For any given strategy profile $\bdelta$ of the leaders or the sellers (see \eqref{eqn_revenue_function}), the optimization problem  ${\cal P}_F (\bdelta, \p)$ of the follower  platform is: 
\begin{equation}\label{eqn_optimization_prob_OS_Game}
\begin{aligned}
\mbox{{\bf   ${\cal P}_F  (\bdelta, \p)$: }}
&\max_{\pi} R(\pi; \bdelta, \p),   \ \text{ subject to } \mbox{an appropriate ordering policy}~\pi.
\end{aligned}    
\end{equation}
As mentioned above, one can obtain the unique `fair' optimizer now in terms (only) of random menus $\Pi^*(\bdelta)$. The utility of seller $\rev{a \in \I}$ equals   (see~\eqref{Eqn_OP_game} and also observe that the prices are set a priori): 
\begin{eqnarray}
\label{Eqn_OS_game}
\U_a (\bdelta):=    \sum_{\pi}  \Pi^*(\bdelta) (\pi)  \sum_{k=1}^M \left (  \indic{ a =  \pi(k)} 
( (1 - \delta_{a}) p^*_{a}  - c_{m_a} ) \beta_{a}(p^*_a) \prod_{j < k} \left(1 - \beta_{\pi(j)}(p^*_{\pi(j)} \right)\gamma  \right ).
\end{eqnarray}

Having defined both the games~\eqref{Eqn_OP_game} and \eqref{Eqn_OS_game}, we begin their analysis by first solving the two lower-level platform optimization problems. We begin with the OP game. 

\ignore{
\begin{lemma}
\label{lemma_joint_opt}
The optimization problem $\max_{\z} R(\z)$ can be simplified as
$$\max_{\z} R(\z) = \max_{\pi} \max_{p} R(\pi, \p),$$
where the function $R$ is defined in \eqref{eqn_revenue_function}.
\end{lemma}

\begin{proof}
Consider any $\lx \subseteq \I$. For any $ \pi \in \Pi(\lx)$ and $\p,$ define
$$
 R(\pi,\p^*) := \max_\p R(\pi,\p). 
$$
Observe that $\Pi(\lx)$ is finite for any finite set $\lx$, price vector $\p$ is defined on a compact set, and $R$ is continuous for every $\pi \in \Pi(\lx)$. Thus, for any $\pi \in \Pi(\lx)$, value of  $R(\pi,\p^*)$ is finite. Thus,
\begin{align*}
   \max_{\z} R(\z) = \max_{\pi,\p} R(\pi, \p) &= \max_{\pi \in \Pi(\lx)} \{R(\pi,\p^*)\}\\
    &= \max_{\pi} \max_{p} R(\pi, \p).
\end{align*}
\end{proof}
}

\section{Platform optimization:  OP game}
\label{sec_MDP}

To begin with, we will show that the platform-optimization problem \eqref{eqn_optimization_prob_RS} can be modeled as an MDP problem. The main idea is to map the item location index to the time index in the MDP formulation.  We begin with the mapping details.

\subsection{MDP formulation}\label{subsec:MDP_Formulation}

Say the items from a  list $\{a_1, a_2, \dots, a_{k-1}\}$ have already been placed in the top $(k-1)$ positions. In this case, the set of available options at location $k$ is $\I - \{a_1, a_2, \dots, a_{k-1}\}$. These options are relevant under two conditions: (a) if the customer is not satisfied with the top $(k-1)$ items, and (b) if the customer is willing to explore more items. If the customer exits the system, it reaches an absorbing state where no further revenue can be generated from this customer. Given this, we can model the remaining items or an absorption-indicator as a state in the MDP being constructed. Let 
$$
\X = \{\lx : \lx \subseteq \I  \mbox{ or } \lx = \Delta \}, 
$$
denote the states of the MDP, where   $\lx = \Delta$ represents an absorbing state. At any position/location index (say $t$), for any state of remaining items~$\lx$ (not the absorption state), one needs to choose an item $a$ from $\lx$ to display on $t$-th position and also decide a price $p \in [0, p^{\text{max}}]$ to quote for that item. 
Thus, the set of actions available at any position-index are dependent on state $\lx$ and the available action space can be denoted by, 
$$
\A_\lx = \{(a,p): a \in \lx, p \in [0, p^{\text{max}}]\}.
$$
The next step is to describe the state `transitions'. Starting from any state $\lx$ and taking an action $(a, p)$, the customer purchases item $a$ with probability $\beta_a(p)$. If the customer does not purchase the item (with probability $1 - \beta_a(p)$) and decides not to continue scanning the remaining items, they exit the system with probability $1 - \gamma$. In either case, whether the customer purchases the item or exits the system, the transition leads to the state $\Delta$.
If the customer does not purchase the item (with probability $1 - \beta_a(p)$) but chooses to continue scanning (with probability $\gamma$), the system transitions to state $\lx' = \lx - \{a\}$.
Thus, the transition probabilities can be summarized as:
$$
\prob(\lx'|\lx, (a,p)) = \begin{cases}
1 & \text{if }  \lx = \lx' = \Delta \\
\gamma (1 - \beta_a(p)) & \text{if } \lx' = \lx - \{a\} ,\\
1 - \gamma (1  - \beta_a(p) )
& \text{if } \lx' = \Delta.
\end{cases}
$$
At any state $\lx,$ with an action $(a,p)$, the platform receives a reward in terms of commission $\delta_a p$, if the customer purchases the product, and zero reward otherwise. Thus, the instantaneous reward function is given by:
$$
r(\lx, (a,p)) = 
\delta_a \beta_a(p) p  \indic{\lx \neq \Delta} .
$$
Having defined the immediate rewards, we next define a policy. A policy dictates the sequence of actions the platform will take throughout the customer's interaction.
In this framework, $\mu = (d_1, d_2, \ldots, d_M)$, with $d_t = \{d_t(\lx)\}_\lx$  for each $t$, represents a policy, where decision rule $d_t(\lx)$ determines the action pair $(a, p)$, given that $\lx$ items are available at step~$t$.  For a Markov deterministic  policy, $d_t(\lx)$ prescribes a fixed $(a,p),$  where
$a$ is the item to be displayed in the $t$-th position, and $p$ is the price quoted for that item. \textit{If $\mu$ is a randomized policy, then $d_t(\lx)$ is  a probability measure on the action-set~$ \A_\lx$, that chooses random actions.} 

Interaction with each customer represents a sample run of MDP, which evolves randomly, running at maximum up to $M$ stages. 
Let $X_t$ denote the set of items available, and $S_t = (A_t, P_t)$ represent the action chosen at step $t$.
The total expected reward under policy $\mu$, starting with the initial set of items $\bar{\lx}$, is given by:
\begin{equation} \label{eqn_MDP_optimization_prob}
    J(\mu, \bar \lx) = \mathbb{E}^\mu \left[ \sum_{t = 1}^M r(X_t, (A_t, P_t)) \right],
\end{equation}
where the expectation $\mathbb{E}^\mu$ is taken accounting for the randomness in customer behaviour and the actions prescribed by the policy $\mu$.
As already mentioned,
the platform can adopt two strategies: a fixed menu, where a complete permutation of items with prices is displayed in one shot, and a randomized menu, where items are offered sequentially with subsequent items chosen probabilistically based on the remaining items\footnote{When the optimizer is not unique for a given state, the fair policy \eqref{Eqn_A_star_tx}-\eqref{Eqn_opt_policy} provided in Section~\ref{subsec:DP_equations} ensures that all sellers (optimizers) are treated fairly and equally at each step. However this can lead to a randomized policy;  in particular by structure of the problem, the optimal policy would suggest one among some finitely many menus in a probabilistic manner; nonetheless the implementation of such a random menu would only mean each customer still gets to see one fixed menu chosen in an IID manner,  implementation details are in  Subsection~\ref{subsec:implementation_of_policy}.
}.
To begin with,  we will
 show that the MDP based cost~\eqref{eqn_MDP_optimization_prob} and that in the  original problem~\eqref{eqn_optimization_prob_RS} are both optimized by a deterministic policy, which is equivalent to a fixed menu-price pair (proof in~\ref{Appendix_for_sec_MDP}):
\begin{lemma}[{\bf Existence}]\label{lem_equivalence_between_random_deterministic_policy}
\begin{enumerate}[(i)] 
    \item For any deterministic policy $\mu,$ there exists a fixed menu-price vector, $\bar \z^\mu = (\pi, \p)$ such that  $ J(\mu, \bar \lx)  = R(\bar \z^\mu, \bar \lx) $, with $R$ defined as in \eqref{eqn_revenue_function}.
    \item Given any $\bar \z = (\pi, \p)$, there exists a policy $\mu_{\bar \z}$ such that 
    $ R(\bar \z, \bar \lx) = J(\mu_{\bar \z}, \bar \lx). $ 
    \item Finally, we have the existence of solution for \eqref{eqn_optimization_prob_RS}  and the MDP for any $\bdelta$:
    $$
    \max_{{\bar \z} = (\pi, \p)} R(\bar \z, \bar \lx) \stackrel{a}{=} \max_{\mu, \mbox{ deterministic}}J( \mu,  \bar \lx) .
    $$
 
\end{enumerate}
\end{lemma}
\noindent Hence, we have the existence of solution for problem  \eqref{eqn_optimization_prob_RS}; we also established that the optimal policy by solving the MDP, or an optimal menu obtained by direct optimization, perform equally well for the platform.  Interestingly, a Markov deterministic policy   is  optimal in this context (see equality $a$ in part~(iii)). However, from a fairness perspective regarding different sellers, a randomized `fair' policy would be preferred (see \eqref{Eqn_A_star_tx}-\eqref{Eqn_opt_policy} in next section). We will discuss this in detail and provide a characterization of both the optimal and `fair' policies in the following section. Additionally, we will show that the performance of the optimal policy, whether derived from a fixed menu or a randomized menu-price strategy, is the same for the platform (as proven in Lemma~\ref{lemma_DP} in the next sub-section).

\subsection{Dynamic programming equations and the optimal-fair policy}
\label{subsec:DP_equations}
Lemma \ref{lem_equivalence_between_random_deterministic_policy} enables us to leverage upon the DP equations to solve the problem \eqref{eqn_optimization_prob_RS} effectively.
Towards this, let $v_t(\lx)$ represent the value function at step~$t$, which is the maximum expected reward achievable from step~$t$ to the final step $M$, given that the set of available items at step $t$ is $\lx$. Formally,
\begin{eqnarray*}
  v_t (\lx) = \sup_{ \mu_t = (d_t, \cdots, d_M)  }
  J_t (\mu_t,  \lx), \mbox{ with, }  J_t (\mu_t,  \lx)  = \mathbb{E}^{\mu_t} \left[ \sum_{k = t}^M r(X_k, (\A_k, \p_k)) | X_t = \lx \right],
\end{eqnarray*}
where $\mu_t = (d_t, \cdots, d_M)$ denotes the policy from step $t$ to step $M$, and $ J_t (\mu_t,  \lx)$ represents the expected reward obtained under this policy.  
Consider the boundary conditions for the value function as follows: when $\lx = \emptyset$, i.e., without any leftover items, the value function is zero. Also, when $\lx = \Delta,$ for all $t$, we define $v_t(\Delta) = 0$. At the final step $M$, for any $\lx,$ the value function is simply the maximum immediate reward from selecting an item $a$ and its price $p$, i.e.,
$$
v_M(\lx) = \max_{a \in \lx, p \in [0, p^{\text{max}}]} \delta_a \beta_a(p) p.
$$
For intermediate steps $t < M$, the value function satisfies the following recursive DP equation for each $\lx$ (see~\citep{puterman}): 
\begin{align}
v_t(\lx) = \max_{a \in \lx, p \in [0, p^{\text{max}}]} \big( r(\lx, (a, p)) + \gamma (1 - \beta_a(p)) v_{t+1}(\lx - \{a\}) \big),\ \ r(\lx, (a,p)) = \delta_a \beta_a(p) p.
\label{eqn_DP_equations}
\end{align}
Clearly, the domain of optimization for each $(t, \lx)$  is compact and the instantaneous reward function $p \mapsto r(\lx, (a, p))$ is continuous, for each $(\lx,a)$. Hence, by \cite[Chapter~4]{puterman}, there exists a solution to the DP equation \eqref{eqn_DP_equations}. Consequently, there exists a solution among the deterministic policies, which serves as the second proof for the existence of a solution to \eqref{eqn_optimization_prob_RS}, this time using MDP theory (recall that existence was alredy established in Lemma~\ref{lem_equivalence_between_random_deterministic_policy}). 
However, more interestingly, the existence of solution for DP equation \eqref{eqn_DP_equations}  also paves way for choosing an unique `fair' policy among  all possible  optimal policy(ies) as explained in the immediate next.

Define the set of optimizers of the equation \eqref{eqn_DP_equations} for each $(t, \lx)$ as follows:
\begin{eqnarray}
\label{Eqn_A_star_tx}
    A^*(t,\lx)  := \Argmax_{a \in \lx, p \in [0, p^{\text{max}}]} \big( \delta_a \beta_a(p) p + \gamma (1 - \beta_a(p)) v_{t+1}(\lx - \{a\}) \big). 
\end{eqnarray}
The above set of optimizers need not be unique for every $(t, \lx)$, however,   the following `fair' randomized policy is one among the optimizers, as established in \cite[Chapter~4]{puterman}:
\begin{eqnarray}
    \z^* (\bdelta):=\{\z^*(t, \lx)\}_{\lx, t} , \mbox{ where  each } \z^*(t, \lx) \mbox{ is uniformly distributed  over $A^*(t,\lx)$} . \label{Eqn_opt_policy}
\end{eqnarray}
This policy ensures that equal preference is given to each of the `equal' sellers at any  $(t,\lx)$, which takes care of fairness.%
\rev{
The problem of fair allocations dates back to several decades ago: for example, in the context of  wireless allocations, authors in \cite{kushner2004convergence} have discussed about proportional fair allocation of channels to various competing users. There were several other papers, hence after,  that discuss fair allocations.  Importance is given to such allocations, as these directly reflect the quality of service (QoS), which in turn reflects the customer satisfaction. 
In our context, fair-allocations imply ordering the `equal' sellers in an `equal' manner. 
When the platform implements \eqref{Eqn_opt_policy}, on average the visibility of all the `equal' sellers would be the same. 
Once again such allocations reflect better QoS to sellers, which in turn can improve the credibility of the platform (thereby improving the interest across sellers). 
}

We now formally summarize the optimality of the fair-policy \eqref{Eqn_opt_policy} below:

\begin{lemma} [{\bf Fair policy}]
 \label{lemma_DP}
There exist a solution to DP equations~\eqref{eqn_DP_equations}. Hence, 
      $$
    \max_{{\bar \z} = (\pi, \p)} R(\bar \z, \bar \lx) = \max_{\mu, \mbox{ deterministic}}J( \mu,  \bar \lx) = \max_{\mu, \mbox{ randomized}}J( \mu,  \bar \lx).
    $$
Thus, the fair-policy \eqref{Eqn_A_star_tx}-\eqref{Eqn_opt_policy} also forms a 
 solution   for  $P_F(\bdelta)$ defined in~\eqref{eqn_optimization_prob_RS} for each $\bdelta$.

 \end{lemma}

Further,  by Assumption~\ref{assumption_beta_fun}, 
we have  unique optimizer $p^*(a,c)$ for each $\mathbb{P}(a,c)$ problem defined below: 
\begin{equation*}
  \mbox{{\bf Problem} } 
  {\mathbb P} (a, c):   \max_p \big( \delta_a \beta_a(p) p + \gamma (1 - \beta_a(p)) c \big). 
\end{equation*}
Thus, $|A^*(t,\lx)| \le  |\lx|$, so the set of optimizers 
in~\eqref{Eqn_A_star_tx} is finite.
Therefore,   the fair policy \eqref{Eqn_A_star_tx}-\eqref{Eqn_opt_policy} can be interpreted as a policy that randomly chooses one among some finitely many permutations $\{\pi_i\}_{i \le l}$  (for some $l < \infty$) to be chosen with 
probabilities  $\{\Pi^*(\pi_i)\}_l$ that are defined using the uniform distributions of \eqref{Eqn_opt_policy}.
For every given $\pi_i$, one can compute the corresponding price-vector  $\{ p_{\pi_i, k}\}_{  k \le M}$ again using the optimizer sets $A^*(t,x)$ of \eqref{Eqn_A_star_tx}, more details are given in equations \eqref{eqn_opt_p_star_last_stage}-\eqref{eqn_price_component_defn} of next section.  
The resultant policy  is exactly 
as in Definition~\ref{defn_policy}.  
We conclude this section with an illustrative example.

\begin{example}[Illustrative example 
with $N=3$ and $M=2$]
\label{ex_illustrative_example}
Consider a setting with $N = 3$ items and a menu of length $M = 2$. At stage 1, we start in the initial state where all items are available: $\I = \{1, 2, 3\}$. In the DP equation \eqref{eqn_DP_equations} corresponding to stage~1, suppose both $(1, p^*_{t=1,1})$ and  $(2, p^*_{t=1, 2})$ are optimizers. Under the fair policy, seller 1 is placed in the top position of the menu with probability $\nicefrac{1}{2}$, and seller 2 is placed at the top with the remaining $\nicefrac{1}{2}$ probability.

\noindent
\begin{minipage}[h]{0.58\textwidth}
\vspace{-3cm}
Accordingly, if the customer does not purchase the first displayed item and chooses to continue browsing, the system transitions to stage 2 with one of two possible states: either $\lx_2 = \{2, 3\}$ (if item 2 was rejected) or $\lx_2 = \{1, 3\}$ (if item 1 was rejected). Suppose the optimal policy in stage 2 recommends: either placing seller 3 at the second position with price $p^*_{{t=2}, 3}$ when state $\lx_2 = \{2, 3\}$, 
or  placing seller 1 at the second position with price $p^*_{{t=2}, 1}$ when state $\lx_2 = \{1, 3\}$.
Thus, the fair optimal policy \eqref{Eqn_A_star_tx}–\eqref{Eqn_opt_policy} randomizes equally between the two permutations, $\pi_1 = (1, 3)$ and $\pi_2 = (2, 1)$, with $\Pi^*(\pi_1) = \Pi^*(\pi_2) = \nicefrac{1}{2}$, as shown in Figure~\ref{fig:illustrative_example_RS}.
 The associated price vector $[p^*_{\pi_1,1}, p^*_{\pi_2, 2}, p^*_{\pi_1, 3}, p^*_{\pi_2, 1}]$ is given by:
 $$
 p^*_{\pi_1,1} =   p^*_{t = 1,1}, \  \ p^*_{\pi_2, 2} = p^*_{t = 1, 2},  \  \ p^*_{\pi_1, 3} = p^*_{t = 2, 3}, \ \ p^*_{\pi_2, 1} = p^*_{t = 2, 1} .
 $$
\end{minipage}%
\hfill
\begin{minipage}[t]{0.38\textwidth}
\centering
\includegraphics[page = 2,trim={6.4cm 1.6cm 6.4cm 0cm}, clip, scale=0.27]{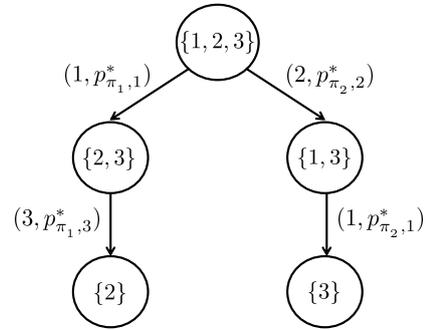}
\captionof{figure}{Example~\ref{ex_illustrative_example}: nodes denote states and edges denote actions, $\Pi^*(\pi_1) = \Pi^*(\pi_2) = \frac{1}{2}$ with $\pi_1 = (1,3)$, $\pi_2 = (2,1)$.}
\label{fig:illustrative_example_RS}
\end{minipage}
\medskip

Besides the fair policy, there also exist two deterministic policies that yield the same revenue for the platform: the first deterministic policy is given by  fixed menu $ \pi_1 = (1, 3)$ and the fixed price vector $[p^*_{\pi_1,1},  p^*_{\pi_1, 3}] $; while,  the second optimal deterministic policy  advocates fixed  menu $\pi_2 = (2, 1)$  with  prices $[p^*_{\pi_2, 2}, p^*_{\pi_2, 1}]$.
All three policies perform equally well from the platform’s perspective. However, the fair policy is preferable from the sellers’ standpoint, and hence we adopt it.
\end{example}

\subsection{Implementation of  fair policy}
\label{subsec:implementation_of_policy}
As described in~\eqref{Eqn_A_star_tx}, when the optimizer is not unique for a given state $(t, \lx)$, we select the fair optimizer defined in~\eqref{Eqn_opt_policy}. This fair selection ensures equal preference among all \textit{equal} sellers at any step. Importantly, such a  fair policy can be computed  offline by solving the DP equations~\eqref{Eqn_A_star_tx} and~\eqref{Eqn_opt_policy}.

The policy might be randomized, nonetheless each customer is presented one fixed menu and the corresponding price vector. Basically, as already explained, by Assumption~\ref{assumption_beta_fun} the random policy supports finitely many menus and the corresponding price-vectors. The offline computation provides us the set of these finitely many options along with their respective probabilities (recall the probability distribution is represented by $\Pi^*$). Thus for each incoming customer, the platform draws one among the menus according to distribution $\Pi^*$, independent of the previous choices. \textit{In all, each customer sees a fixed menu, but the randomization ensures fairness among sellers. } 
For instance, in Example~\ref{ex_illustrative_example}, the platform presents the customers either menu $(1,3)$ or menu $(2,1)$ with equal probabilities.

\section {Small-$\gamma$ approximation for OP game}
\label{sec_small_gamma_game}
In this section, we derive the precise details (closed-form expression) of the platform's optimal policy in the asymptotic regime, where the continuation probability $\gamma$ is small.  

\subsection*{Optimizer of follower in OP game for a special case } 
We obtain the optimal and fair (or $ \epsilon $-optimal) policy, as given in \eqref{Eqn_A_star_tx}--\eqref{Eqn_opt_policy}, by solving the DP equations~\eqref{eqn_DP_equations} with the initial state $ \bar \lx = \I $ and for small $ \gamma $. This is achieved by first constructing a randomized policy and then proving that it satisfies \eqref{eqn_DP_equations}. We begin by defining an index function $ \fstar(a) $, which captures the maximum immediate reward obtainable by the platform from the seller-item $ a $:

\begin{equation} \label{eqn_f_star_index}
\fstar(a):= \max_{p} \delta_a \beta_a(p) p.
\end{equation}

Towards constructing the policy, for any subset $ \lx \subset \I $, define the permutation $ (a_1(\lx), \dots, a_{|\lx|}(\lx)) $ as the elements of $ \lx $ sorted in non-increasing order of their $ \fstar $ values, i.e.,
\begin{equation} \label{eqn_f_star_inequality}
\mbox{if }\ {i \le j} \ \text{ then }\ \fstar(a_i (\lx) ) \ge \fstar ( a_j (\lx) ).   
\end{equation}
If ties exist among items with the same $\fstar$ values, i.e., if {\small$\fstar(a_i(\lx)) = \cdots = \fstar(a_{i+j}(\lx))$} for some $i$ and $j \ge 1$, then these $(j+1)$ items can be placed  uniformly (random)   among positions~$i$ to $i+j$. If $i + j > M,$ then in every possible random menu of length $M$, exactly {\small$i + j - M$} of these items do not get a position, and this selection is again uniformly distributed.

Let $\pi^* := (a_i(\I): i \leq M)$ be a permutation of top $M$ ordered elements,  picked according to $\fstar$-index   and with uniform (random) selection for `equal'-$\fstar$ sellers, as above:  
\begin{align}
 \pi^* &  = (a_1 (\I), a_2 (\I), \cdots, a_{M} (\I) ), \mbox{ with, }  \Pi^*(\pi^*) := \prob(\mbox{menu} = \pi^*).
 \label{eqn_opt_perm_small_gamma}
\end{align}
Basically, $\Pi^*(\pi^*)$ is the   probability of picking $\pi^*$ via `uniform' selection. One can easily compute the above probability $\Pi^*$ (see the details  of deriving  \eqref{eqn_unified_seller_utility} of  the next section and see also the Example~\ref{ex_illustrative_example}). 
For each   $\pi^*$ (with $\Pi^*(\pi^*) > 0$), we now define a price vector $\p^\sstar (\pi^*) =( p_k^\sstar(\pi^*))_k$, in backward recursive manner.
For the last step  or the last position in menu ($k = M$), the optimal price (depends on $\pi^*$, \eqref{eqn_opt_perm_small_gamma}) is:  
\begin{equation}\label{eqn_opt_p_star_last_stage}
p_M^\sstar (\pi^*) \in \argmax_{p \in [0, p^{\text{max}}]} ( \delta_a \beta_a(p) p), \mbox{ where } a = a_M(\I), \mbox{ and } \ \ v_M^\sstar := \fstar(a_M(\I)).
\end{equation} 
For a position $k < M$ in the menu, the optimal price $p_k^\sstar(\pi^*)$, which is unique under the Assumption \ref{assumption_beta_fun}, is given by the following expression, where $a = a_k(\I)$ of permutation $\pi^*$:
\begin{eqnarray}\label{eqn_opt_p_star}
p_k^\sstar (\pi^*) \in \argmax_{p \in [0, p^{\text{max}}]} \left\{ \beta_{a}(p) \left(p \delta_{\small a } - \gamma v_{k+1}^\sstar\right) \right\}, \mbox{ and } v_k^\sstar = \beta_{a }(p_k^\sstar(\pi^*) ) \left(p_k^\sstar (\pi^*) \delta_{a } - \gamma v_{k+1}^\sstar\right) + \gamma v_{k+1}^\sstar.     
\end{eqnarray}  
For a given menu $\pi^*$, the above price vector is position-dependent. We convert it into a seller-dependent version as below, which provides  the potential price-component of $\z^*(\bdelta)$ corresponding to $\pi^*$:
\begin{eqnarray}\label{eqn_price_component_defn}
    p_{\pi^*, a}^* := \sum_{k} p_k^\sstar (\pi^*) \indic{ a =  a_k(\I)   } , \mbox{ for each }  a. 
\end{eqnarray}
Following a similar procedure, one can define $\{p^*_{\pi, a}\}_a$ for all $\pi$ with $\Pi^*(\pi) >0.$
 
\textit{We now prove that $\z^*=\z^*(\bdelta) := (\Pi^*, \{p^*_{\pi,a}\}_{\pi, a})$  is the fair and  optimal policy, as given by~\eqref{Eqn_A_star_tx}–\eqref{Eqn_opt_policy}, for  small enough~$\gamma $, when there are no `equal' sellers}  (proof in  \ref{sec_Appendix_sec_four}).

\begin{theorem}
\label{thm_small_gamma_optimality}
Consider any $\beta_a(\cdot)$ function satisfying Assumption~\ref{assumption_beta_fun}. Further, assume that the $\fstar$-values are strictly decreasing, i.e., $\fstar(a_k(\I)) > \fstar(a_{k+1}(\I))$ for all $k$. Then, there exists a threshold $\bar \gamma > 0$ such that for all $\gamma < \bar \gamma$, the fair policy $\zsstar(\bdelta)$ is the only optimal policy.
\end{theorem}

\subsection*{Approximation for general case} 
Under the hypothesis of Theorem~\ref{thm_small_gamma_optimality}, for small values of $\gamma$, the Markov randomized (fair) policy given by \eqref{Eqn_A_star_tx}–\eqref{Eqn_opt_policy} simplifies to \eqref{eqn_opt_perm_small_gamma} when the initial state is $\bar \lx = \I$ (we have a deterministic policy here, i.e., $\Pi^*(\pi^*) = 1$ for the unique $\pi^*$   defined in \eqref{eqn_opt_perm_small_gamma}). In other words, for sufficiently small $\gamma$ and with `non-equal' sellers, the (fair) policy always suggests $a_t (\I)$ at step $t$, whereas for larger $\gamma$, the decision may depend on $\lx$ as shown in~\eqref{Eqn_opt_policy}.

This result provides a valuable structural insight: when the continuation probability is sufficiently small (i.e., when $\gamma$ is small), the optimal strategy simplifies to a greedy, myopic selection rule that prioritizes items solely based on their immediate reward potential, as indicated by the index rule $\fstar$ in \eqref{eqn_f_star_index}; this is true for case where `equal' sellers are absent (observe strict inequalities in the hypothesis of Theorem \ref{thm_small_gamma_optimality} for $f^*$ indices).   

We now identify even simpler policies that are $\epsilon$-optimal, and, in fact, apply to a broader range of scenarios than those covered in Theorem~\ref{thm_small_gamma_optimality}:
\begin{enumerate}[(i)]

\item It is important to observe here (and prove, in a manner similar to Theorem~\ref{thm_small_gamma_optimality}) that if the strict inequality assumption $\fstar(a_k(\I)) > \fstar(a_{k+1}(\I))$ is relaxed to $\fstar(a_k(\I)) \geq \fstar(a_{k+1}(\I))$, the policy $\zsstar$ from Theorem~\ref{thm_small_gamma_optimality} remains $\epsilon$-optimal --- this step is omitted here, as the proof can be adapted easily. In other words, under the hypothesis of Theorem~\ref{thm_small_gamma_optimality}, $\Pi^*(\pi^*) = 1$ for a fixed menu $\pi^*$, while after relaxing the assumption, we will have randomization, with $\Pi^*(\pi) \in (0,1)$ for some $\{\pi\}$. 

\item  For small values of $\gamma$, one can simplify further by replacing $\{p^*_{\pi,a} \}$ of $\z^*(\bdelta)$ given in \eqref{eqn_price_component_defn} with $\{p^*_a\}$ defined in~\eqref{eqn_item_prices_OS_game}. We refer to such  a policy   as $\z^\epsilon(\bdelta)$. One  can easily  prove  that  $\z^\epsilon(\bdelta)$ is $\epsilon$-optimal as  shown in  Theorem \ref{thm_small_gamma_optimality}, and this can be verified by observing  \eqref{eqn_opt_p_star_last_stage}-\eqref{eqn_opt_p_star} for small~$\gamma$. In fact, $\z^\epsilon(\delta)$ is also $\epsilon$-optimal for OS game, which we will prove next.  Formally, $\z^\epsilon (\bdelta)$ is given by:
\begin{equation}
\label{eqn_small_gam_approx_policy}
\begin{aligned}
\hspace{-1.3cm}\mbox{\bf Small-$\gamma$ approximate policy }  \z^\epsilon (\bdelta): \quad \ \ & \\     p^\epsilon_{\pi, a} = p^*_a = \arg \max_p & (\beta_a(p) p)  \mbox{ for each } a \mbox{ as in \eqref{eqn_item_prices_OS_game}},  \hspace{0mm}  \mbox{ and }  \Pi^\epsilon = \Pi^* \mbox{  of \eqref{eqn_opt_perm_small_gamma}}.
\end{aligned}
\end{equation}
\end{enumerate}

\section{Platform optimization:   OS game and small $\gamma$-approximation} 
\label{sec_small_OS}
The platform's optimization problem for the OS game~\eqref{eqn_optimization_prob_OS_Game} has already been addressed in the literature. For example, in \cite{dhingra2021multi, kempe2008cascade}, an index policy called the `$h$-index' is proposed, which provides the optimal menu order that solves~\eqref{eqn_optimization_prob_OS_Game}. The $h$-index for seller $a$ is defined using the price vector $\{p^*_a\}$ from \eqref{eqn_item_prices_OS_game}: 
\begin{equation}\label{eqn_h_index}
h(a, p_a^*) = \frac{\beta_a(p_a^*) p_a^* \delta_a^*}{1 - \gamma (1 - \beta_a(p_a^*))},  
\end{equation}  
and the platform ranks sellers in the  decreasing (more precisely, in non-increasing) order of their $h$-index values. 
We now state the following Lemma, which directly follows from \cite[Theorem~2]{dhingra2021multi} and the results in \cite{kempe2008cascade}. Furthermore, the proof of $\epsilon$-optimality or optimality at small $\gamma$ is straightforward from \eqref{eqn_h_index}.

\begin{lemma} \label{lem_f_index_optimality}
\begin{enumerate}[(i)]
    \item Assume   $\beta_a(p_a^*) p_a^* \delta_a^* \neq \beta_b(p_b^*) p_b^* \delta_b^*$ for all $a, b \in \I$. Then, there exist  a threshold $\bar \gamma > 0$ such that for all $\gamma < \bar \gamma$, the menu obtained by ordering according to the $h$-index values, i.e., the optimal menu for problem \eqref{eqn_optimization_prob_OS_Game},  is the same as that obtained by ordering according to   $\fstar$ indices defined in~\eqref{eqn_f_star_index}. 
    
    \item If $\beta_a(p_a^*) p_a^* \delta_a^* = \beta_b(p_b^*) p_b^* \delta_b^*$ for some $a, b \in \I$, then selecting the menus  randomly  according to  $\fstar$ indices as in \eqref{eqn_opt_perm_small_gamma} is an $\epsilon$-optimal policy.

\end{enumerate}
\end{lemma}

Interestingly, the optimal menu selection for the OP game, as provided by Theorem~\ref{thm_small_gamma_optimality} and in~\eqref{eqn_small_gam_approx_policy}, coincides with that for the OS game. The above lemma demonstrates that the $\fstar$-index and the $h$-index coincide for small values of $\gamma$ (or low customer exploration probability). Furthermore, the optimal (more precisely, $\epsilon$-optimal) prices chosen by the platform in the OP game, as in~\eqref{eqn_small_gam_approx_policy}, coincide with the prices pre-set by sellers (according to~\eqref{eqn_item_prices_OS_game}) in the OS game.

When approximating the seller-utility functions \eqref{Eqn_OP_game} and \eqref{Eqn_OS_game} for the OP and OS games, respectively, using the small~$\gamma$ approximation as in~\eqref{eqn_small_gam_approx_policy} and Lemma~\ref{lem_f_index_optimality}, the resulting seller-utility functions (of $\bdelta$) are exactly the same (with more details provided in the next section). Hence, the top-level (seller) game in both cases coincides under this approximation. This enables us to analyze both games under a unified framework, which we consider in the subsequent sections.

\section{
Unified seller game under small-$\gamma$ regime} 
\label{sec:game_theoretic_modeling}

In this paper, 
we study two games --- the OP   and OS games. Our goal is to analyze these games using the Stackelberg framework, where the platform acts as the follower and the sellers as the leaders.  In Section~\ref{sec_small_gamma_game}-\ref{sec_small_OS} we 
derived $\epsilon$-optimal policy for the platform for each $\bdelta$  (the vector of commissions set by sellers) and for both the games; recall, the approximation is valid for small values of $\gamma$.  

Now, we are left to analyse the upper-level game: a non-cooperative game among $N$ sellers who strategically choose their commissions, $\bdelta=\{\delta_a\}_a$. In the OP game,   the platform sets the prices and chooses the menu according to the $\epsilon$-optimal policy \eqref{eqn_small_gam_approx_policy}, in response to sellers' actions $\bdelta$.  On the other hand,  in the OS game, the prices are pre-fixed by the sellers themselves at \eqref{eqn_item_prices_OS_game}, while the $\epsilon$-optimal menu choice coincides with that in \eqref{eqn_small_gam_approx_policy}; hence, as already mentioned, both the top-level games coincide, under this approximation. In this section, we obtain the closed-form expression for this common utility function of the sellers. In the next section,   the resultant (common) non-cooperative seller  game,  henceforth referred to  as `\ospgame', is analyzed.

We now consider some specific price sensitivity functions which are widely used in literature to proceed further with the analysis. In particular,  the  following exponential price sensitivity functions are considered (e.g., \cite{rader1973nice_demand_function, huang2013demand_all, demand_f_1, talluri2006theory_first_expo_demand, ozer2012oxford_second_expo_demand, Gallego2019_expo_demand_fun, koffarnus2015modified_expo_demand_second_last, hursh2008economic_last_expo_demand}):
\begin{equation}
\label{Eqn_exp_beta}
\beta_a(p) = e^{-\alpha_a p}, \text{ where $\alpha_a$ is the price sensitivity parameter related to item $a$.}
\end{equation}
This exponential structure reflects the commonly observed behavioural pattern: as price increases, the likelihood of a customer purchasing the item decreases exponentially. The rate at which this probability decays with the price is governed by the price sensitivity parameter $\alpha_a$. Such a form ensures analytical tractability and aligns with the observed customer behaviours in various markets \citep{talluri2006theory_first_expo_demand, ozer2012oxford_second_expo_demand, Gallego2019_expo_demand_fun}.


Next, we derive the closed-form expression for the seller-utility function under policy~\eqref{eqn_small_gam_approx_policy}, and
with the 
 exponential price sensitivity function~\eqref{Eqn_exp_beta} (with slight abuse of reference, we refer~\eqref{eqn_small_gam_approx_policy} as optimal policy). To begin with, the prices  of the seller items under both~\eqref{eqn_item_prices_OS_game} and  \eqref{eqn_small_gam_approx_policy},  are given by:
\begin{equation} \label{eqn_opt_price_and_beta_approx}
    p_a^*(\bdelta)
    = \frac{1}{\alpha_{ a}} \ \text{ for all } a , \ \  \text{ and,  }  \ \ \beta_a(p_a^*(\bdelta)) = e^{-\alpha_a p_a^*(\bdelta)}= e^{-1}. 
\end{equation}
Further, from \eqref{eqn_f_star_index}, we have $\fstar(a) = \nicefrac{\delta_a}{e \alpha_a}$ for each $a$; thus the platform determines the set of $M$ items (or the menu) by ranking sellers according to this $\fstar$ index, or equivalently according the values of $\{\nicefrac{\delta_a}{\alpha_a}\}_a$. Therefore, under policy   \eqref{eqn_small_gam_approx_policy},  given the  strategy profile $\bdelta  =\{\delta_a\}_a$ of sellers,  it chooses one among the following menus/permutations (with equally likely selection at the points of ties, see \eqref{eqn_f_star_inequality}-\eqref{eqn_opt_perm_small_gamma})
\begin{eqnarray*}
     \pi^\epsilon(\bdelta) &=& (a_1^\epsilon(\bdelta), a_2^\epsilon (\bdelta), \cdots, a_M^\epsilon (\bdelta)),  \ \ 
     \mbox{ where }
     a_k^\epsilon (\bdelta)  \in   {\Arg    \max}_{a \notin \{a_1^\epsilon(\bdelta), a_2^\epsilon (\bdelta), \cdots,a_{k-1}^\epsilon(\bdelta)\}} \left( \frac{\delta_a}{\alpha_a} \right) .
\end{eqnarray*}
Further, the probabilities $\{\Pi^*(\pi^\epsilon)\}$ in \eqref{eqn_opt_perm_small_gamma}, should be calculated based on  equally likely  random selection  among the sellers with equal values of  $\nicefrac{\delta_a}{\alpha_a}$, at the tie-positions. We now delve into more precise details of this randomization. Recall, such a randomization is important from fairness perspective, see \eqref{Eqn_opt_policy}.  


For given $\bdelta$, the platform categorises the sellers into bins $\{ \B_i \} = \{ \B_i (\bdelta) \} $ using the order statistics of $\{\fstar(a) \}$ as below:
\begin{equation*}
    \B_1 = \Arg \max_a \fstar(a) ,  \ \B_k = \Argmax_{a \notin \cup_{l< k} \B_{l} } \fstar(a),  \mbox{   for any } k > 1. 
\end{equation*}
Let $ k_l := \sum_{i< l} |\B_i| $ represents the number of items preceding the items of bin $\B_l$ in the optimal menu. Let $\l$ denote the last bin,
which is given by: 
$
\l := \max \{ l : k_{ l } < N \}, \mbox{ and observe }  k_{\l} = N.
$
For ease of notation, let $k_{\l +1} := N$.
For any $\bdelta$, the optimal menu for the Stackelberg follower (the platform) is to place the items of bin $B_l$ between positions $k_l$ to  $k_{l+1}$. Further, the items of the sellers in the same bin are uniformly (randomly) placed in these positions, see~\eqref{eqn_opt_perm_small_gamma}.

Now, we are ready to define the utility of the sellers. Each seller $a$ commits $\delta_a$ fraction of its profit to the platform, and thus, for each product sold, using \eqref{eqn_opt_price_and_beta_approx}, it derives a positive utility $
(1-\delta_a) p_a^* = \nicefrac{ (1-\delta_a)}{\alpha_a}.$
If $a \in \B_l = \B_l(\bdelta)$, then the utility of seller~$a$ in both the games \eqref{Eqn_OP_game} and \eqref{Eqn_OS_game} simplifies as follows: 
\begin{align}
\label{eqn_unified_seller_utility}
\U_a(\delta_a, \delta_{-a}) &=
\left( (1-\delta_a) p_a^* - c_{m_a} \right)
\frac{{\tgam}^{k_l}\sum_{0 \leq i < \b_l} {\tgam}^{i} \indic{k_l+i \le M} }{\b_l} =
\frac{ \left( \eta_a -  \delta_a \right)  }{e \alpha_a} \frac{{\tgam}^{k_l}\sum_{0 \leq i < \b_l} {\tgam}^{i} \indic{k_l+i \le M} }{\b_l},\\
\mbox{with, } \  \eta_a &:= 1 - c_{m_a} \alpha_a, \ \tilde{\gamma} := \gamma (1 - e^{-1}),  \text{ and } \ 
b_l :=  |\B_l(  \bdelta)|.    \nonumber
\end{align}
In the above, $\eta_a$ is representative of the proportion of the profit for the seller, after covering the manufacturing cost~$c_{m_a}$. Basically, the term $\nicefrac{(\eta_a - \delta_a)}{e\alpha_a}$ represents the expected revenue derived by seller $a$, conditioned that the customer explores its item, while the second term represents the probability that the customer explores the item of seller $a$ given that the arrangement of items are according to the (possibly randomized) optimal menu~\eqref{eqn_opt_perm_small_gamma}, \eqref{eqn_small_gam_approx_policy} under~$\bdelta$  (recall here that $e^{-1}$ is the probability that the customer buys the item explored). 
 
\textit{The utility function in \eqref{eqn_unified_seller_utility} represents the \ospgame.}
The analysis of this common game helps us to analyse the top level game in both OS and OP games, thereby in completing the analysis of the respective Stackelberg games. Before proceeding further, we will digress slightly to discuss the accuracy of the approximation.

\subsection{Accuracy of small-$\gamma$ approximation}
To begin with, we will observe that for exponential  response function \eqref{Eqn_exp_beta}, the h-index policy \eqref{eqn_h_index} coincides with the $\fstar$-index policy \eqref{eqn_f_star_index}, in fact for all $\gamma$. 
\begin{lemma}\label{lem_OS_OSP_similarity}
For the exponential response function in~\eqref{Eqn_exp_beta}, the seller rankings induced by the h-index  \eqref{eqn_h_index} and the $\fstar$-index  \eqref{eqn_f_star_index} are identical.
\end{lemma}

In view of the above result (proof in \ref{appendix_sec:game_theoretic_modeling}, for the response function~\eqref{Eqn_exp_beta}, it is clear that the exact optimal policy of the platform coincides with small-$\gamma$ approximate policy for all $\gamma$ for the OS game. \textit{Consequently, the utility of any seller in the OS game~\eqref{Eqn_OS_game}  exactly equals that in \eqref{eqn_unified_seller_utility} and hence the  approximate \ospgame~coincides with the OS game for all $\gamma$} (recall in OS game, the prices are pre-set according to \eqref{eqn_item_prices_OS_game}).
 We now examine numerically the accuracy of the small $\gamma$ approximation for OP game.
\begin{figure}[ht]
\vspace{-4mm}
    \centering
    \subfloat[\centering Seller utility under OP and approximate policy for different values of~$\gamma$]{{\includegraphics[trim = {0cm 7.5cm .8cm 8.5cm}, clip, scale = 0.33]{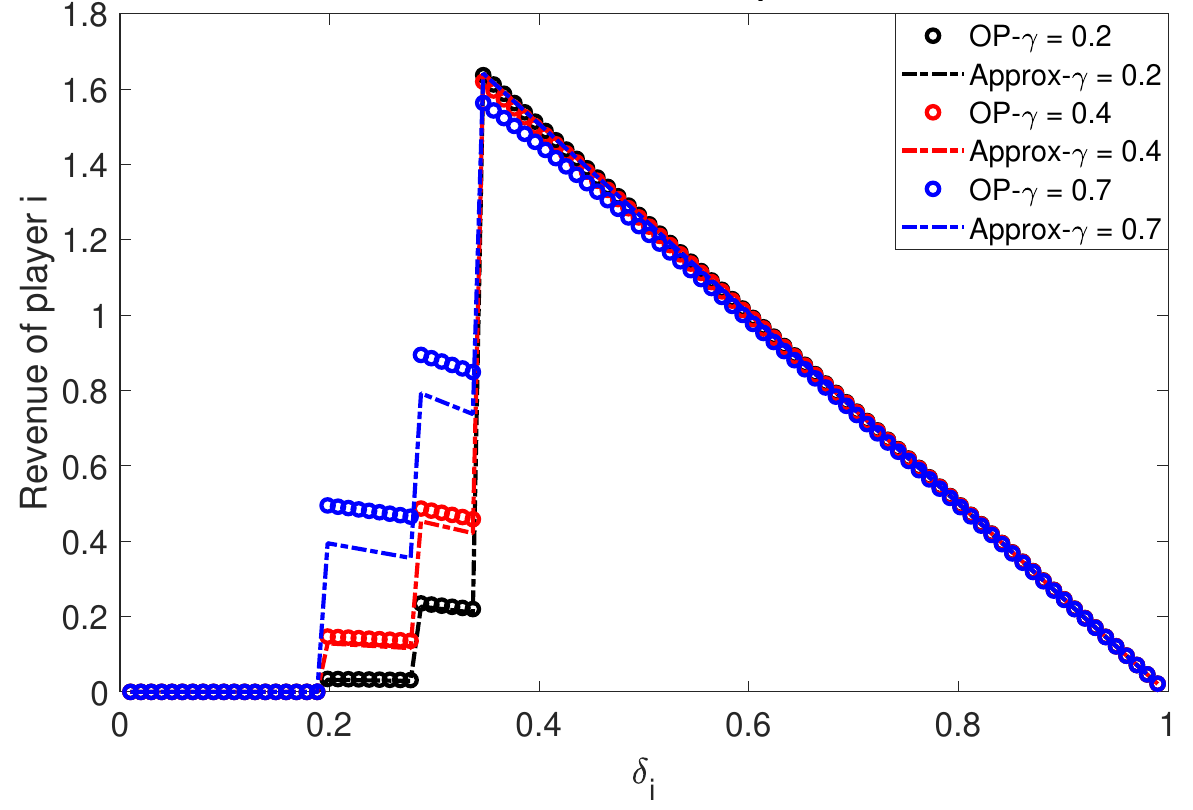}\label{fig:original_utility} }}%
    \hspace{0.6cm}
    \subfloat[\centering Zoomed-in view of Figure~\ref{fig:original_utility} highlighting regions with visible deviations between the curves]{{ \includegraphics[trim = {0cm 7.5cm .8cm 8.5cm}, clip, scale = 0.33]{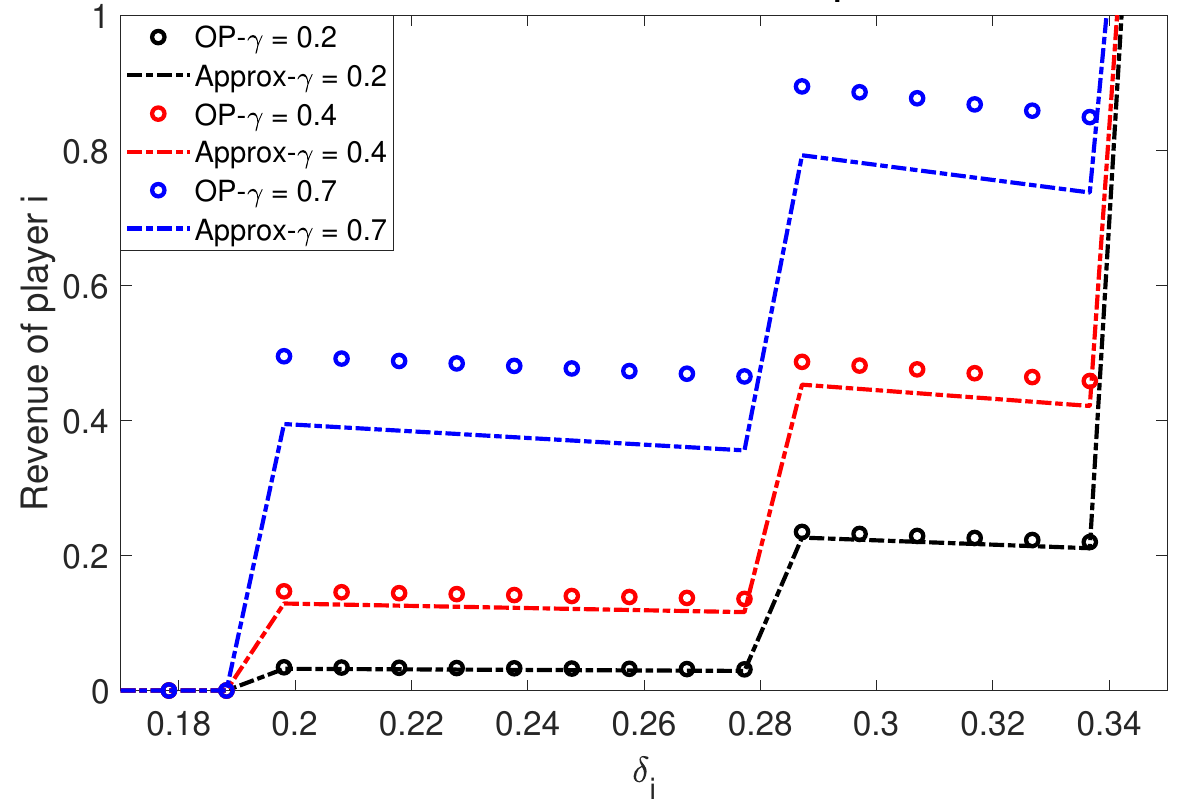}\label{fig:zoomed_utility} }}%
    \caption{Comparison of seller utilities under the OP game~\eqref{Eqn_OP_game} and the approximate policy~\eqref{eqn_opt_perm_small_gamma} for $N=7$ and $M=3$, across different values of $\gamma$.}%
    \label{fig:utility_plot}%
\end{figure}

Towards the above comparison,  we solve the DP equations \eqref{eqn_DP_equations} numerically and estimate the seller utilities at the computed optimal policy; such computations are derived for some selected values of $\delta_a$ of   seller $a$ for some $a$, while $\bdelta_{-a}$ is kept fixed; we also compute the seller utilities under the approximate policy given in~\eqref{Eqn_A_star_tx}–\eqref{Eqn_opt_policy}, for the same values of $\gamma$ and $\bdelta$ and compare the two sets of estimated utilities.  The results are plotted in Figure~\ref{fig:utility_plot}.

All the circles in Figure~\ref{fig:utility_plot} represent the utilities under OP game, which are obtained by solving the DP equations \eqref{eqn_DP_equations} numerically and computing the seller utility functions at the randomised fair policy~\eqref{Eqn_A_star_tx}–\eqref{Eqn_opt_policy}. While, the dotted lines are obtained by computing the same  under approximate policy \eqref{eqn_opt_perm_small_gamma}-\eqref{eqn_small_gam_approx_policy}. One can use the policy evaluation technique to compute the utilities at any given policy (at \eqref{eqn_opt_perm_small_gamma} as well as at the optimal policy obtained by solving the  DP equations, at any  given $\gamma$); we simplify by just considering those values of $\bdelta$ for numerical illustration,  at which both the fair policies  are deterministic.  

Figure~\ref{fig:utility_plot} compares the seller utilities under the OP game and the approximate policy for varying values of~$\gamma$. For $\gamma = 0.2$, the two curves align almost perfectly, indicating almost exact utility matching. As $\gamma$ increases to $0.4$ and $0.7$, the curves start to diverge slightly. To better visualize these deviations, Figure~\ref{fig:zoomed_utility} presents a zoomed-in version of the original plot shown in Figure~\ref{fig:original_utility}, focusing on the regions of $\bdelta$ where the differences are more noticeable: the percentage error is  $1.17\%$, $2.54\%$ and $5.14\%$ respectively  for  $\gamma = 0.2$, $\gamma =0.4$  and $\gamma = 0.7$.  This demonstrates that the approximate policy   provides  a sufficiently accurate estimates of seller utilities even for moderate values of~$\gamma$.
Next, we move to the analysis of the \ospgame.

\section{
Unified seller game: cyclic equilibria}\label{sec:equilibrium}

In this section, we derive the   `solution' of the \ospgame,  which unifies the OS and OP games in small-$\gamma$ regime, using the game-theoretic framework~\eqref{eqn_unified_seller_utility} derived in the previous section. 
Recall by Lemma~\ref{lem_OS_OSP_similarity} that for exponential response function, the OS game coincides with the small-$\gamma$ \ospgame~for all~$\gamma$. Thus, the solutions of the \ospgame, derived in this section, form the solution of the exact OS game with exponential response function, and this is true  for any value of  $\gamma$.

Typically, the solution of a non-cooperative game (corresponding to the top-level in Stackelberg OS/OP framework), is represented by  a Nash equilibrium (NE) -- such equilibrium commission vectors $\bdelta^*$ are stable against unilateral deviations. Interestingly, in  (both) our seller-platform interaction games, we observe the emergence of a cyclic equilibria. Recently, the authors in~\cite{EC_Paper} proposed a solution concept called \textit{equilibrium cycle} for the games without pure strategy NE, and we would prove the existence of such cycles as a solution for our unified game. \textit{Existence of such cyclic equilibria implies that the sellers vary commissions in a cyclic manner at the equilibrium}. This section will have a detailed discussion of these topics.

Further,  from~\eqref{eqn_unified_seller_utility}, it is evident that the utility of   seller $a$ is dictated by  factor $\eta_a = (1-c_{m_a} 
\alpha_a)$. This factor denotes the highest sustainable commission/margin for seller~$a$ --- it would derive negative utility for commissions higher than $\eta_a$. This kind of structure indicates that the share of the seller at any equilibrium/solution of the game depends primarily upon the magnitude of this value~$\eta_a$.  \textit{In this section, we will prove that the set of factors $\{\eta_a\}_a$ are, in fact, the sole determinants of the game's outcome and the equilibrium utilities of the individual sellers.}
Interestingly, there are two independent levers that can determine any sellers' capacity: $c_{m_a}$ the production capacity and $\alpha_a$ the price-sensitivity factor representative of the reputation among the customers.  However, the strategic interactions between the sellers and the platform appear to be driven by a single, consolidated factor, $\eta_a$, which is a specific function of these two `independent' levers~$c_{m_a}$ and~$\alpha_a$. 

We now delve into the analysis of the \ospgame:  we immediately begin 
with almost equal sellers, while  the emergence of cyclic equilibria is discussed in Section~\ref{subsec:unequal_seller_EC}, while studying  the scenarios with  significantly different sellers.

\subsection{Equal sellers: Nash or $\epsilon$-Nash equilibria}

We now consider `near symmetric' environments, 
by considering the cases where the sellers are relatively similar in their reputation or strength, represented by almost equal $\{\eta_a\}_a$ values in \eqref{eqn_unified_seller_utility}.
This 
game admits the typical solution in the form  of a NE or an $\epsilon$-NE. 
The following theorem 
formally characterizes two  types of scenarios and identifies the corresponding  equilibria (proof is provided  in~\ref{appendix_sec:equilibrium}).

\begin{theorem}
\label{thm_equal_sellers_equilibrium}
Assume $N > M$.
\begin{enumerate}[(i)]
    \item  
    Suppose there exists $L \ge M+1$ such that $\eta_k = \eta_c$ for all $k \le L$, and $\eta_k < \eta_c$ for all $k > L$. 
    Then the strategy profile $\bdelta_* = (\eta_1, \eta_2, \cdots, \eta_{L}, \delta_{L+1}, \cdots, \delta_N)$ is a Nash equilibrium, if $\delta_k \le \eta_k$ for each $k > L$.  
  
\item  Suppose the $\{\eta_k\}$ values are as below: \begin{equation} \label{eqn_assump_seller_eta_different}
\eta_1   > \eta_2 > \cdots     > \eta_N >  0.    
\end{equation} 
Further, assume that $(\eta_{1} - \eta_{M+1})  <  \epsilon \min \left  \{ e \min_{k\le M}\alpha_k, 1 \right \}.$
Then the  strategy profile
$$
\bdelta_* = (\eta_2, \eta_3, \cdots, \eta_{M+1}, \eta_{M+1}, \delta_{M+2}, \cdots, \delta_N)
$$ 
is an $\epsilon$-equilibrium, if   $\delta_k \le  \eta_k$ for all {\small $k > (M+1)$}.

\end{enumerate}
\end{theorem}

Thus, by the above theorem, when there are $L$ exactly equal sellers (in terms of $\eta_a$) with  $L \ge M+1$ and when these $L$ sellers outshine the rest, the \ospgame~has many NE, in all of which the `bigger and equal' sellers use common commission $\eta_c$, see part~(i). This resultant \textit{equilibrium is similar to a classical Bertrand duopoly equilibrium, where the competition among the symmetric sellers forces them to operate at the marginal cost} ---  the sellers derive zero utility at NE. As seen in the proof, the top $L$ sellers are placed uniformly in the top $M$ positions of  the menu at the resultant equilibrium; in other words, the items of  any seller are equally likely to be placed  at any   position in  the menus displayed to different customers.  




When the sellers are of distinct  strengths but the best seller (with highest reputation factor) is not significantly stronger than the $(M+1)$-th seller,  we have an $\epsilon$-equilibrium, see  part~(ii). Basically, when the gap $\eta_1 - \eta_{M+1}$ is sufficiently small, there exist an $\epsilon$-equilibrium. Interestingly, seller~$1$ chooses $\eta_2$ as the commission, seller~$2$ chooses $\eta_3$, and in general seller $k$ chooses $\eta_{k+1}$ as the commission at the  equilibrium.
Only the top $M$-sellers obtain a  position in the menu (seller $k$ obtains $k$-th position) at the equilibrium.


\ignore{
\begin{definition}[Better response chain]
    We say $\act \to \act'$, if there exists a player $i$ with $a_i \ne a_i'$,  $a_j = a'_j$ for all $j \ne i$ and
    $$
 \U_i (\act') > \U_i (\act).
    $$
    A set of points $(\act_1, \act_2, \cdots, \act_n)$ is called $n$-hop chain, if for every $l<n$, $\act_l \to \act_{l+1}$. 
\end{definition}}



We have so far focused on the settings where the sellers possess equal or nearly equal strengths, which admit classical pure strategy NE or $\epsilon$-NE. In this regime, the  equilibrium outcomes are analytically tractable and interpretable. However, as we move to the settings with asymmetric sellers, we encounter fundamentally different behaviour. 
When the seller reputations are significantly different, one does not have a classical pure strategy NE and the `best' response dynamics oscillate (see~Section~\ref{sec:numerical_results}). One requires a broader notion  of equilibrium that can capture such oscillations; and we resort to a recent solution concept proposed in~\cite{EC_Paper}. 
The `equilibrium cycle' (EC) developed in~\cite{EC_Paper} is powerful enough to capture  the potential fluctuations that would be persistent even at the equilibrium (or even after a long run)  for games (like ours) without  standard NE; basically such concepts capture the equilibrium patterns in the scenarios without point solutions. 

When we attempt to adapt the framework of~\cite{EC_Paper} for analysis, we encounter a problem on a set of measure zero; it is amply clear in the literature that sets of measure zero 
often do not alter the outcomes or the consequences. In this paper, we manage to extend the notion of EC to navigate through such measure-zero violations and  establish  that our \ospgame~posses the modified EC as a solution. 
We now take a brief detour to explain the framework of \cite{EC_Paper},  and then to extend the notions.

\subsection{$\mu$-Connected Equilibrium Cycle}
Recently, the authors in~\cite{EC_Paper, papadimitriou2019game}, illustrate that dynamics in many strategic-form games may not converge to a fixed point (like NE), but rather oscillate within some finite intervals.
In this section, 
we build upon the idea of EC  proposed in~\cite{EC_Paper} and introduce a more general solution concept, namely the `$\mu$-connected equilibrium cycle' ($\mu$-EC).
To formally define $\mu$-EC, we first define its key components: $\epsilon$-best response chains, $(n, \epsilon)$-tails, and external tails. This construction allows us to \textit{represent how players iteratively adjust their strategies in response to others, and how such sequences of responses encompass some recurring paths of strategic behaviours leading to a `generalized notion of dynamic equilibrium.'}

Consider a strategic form game $ \left< \N, (\A_i)_{i \in \N}, (\U_i)_{i \in \N}\right>,$ where $\N = \{1,\cdots,N\}$, for each~$i$, $\A_i$ is a non-empty set in a metric space.  In many such games, especially in games with discontinuous utility functions, a player may not have a best response against a given strategy profile of opponents -- the players may instead opt for near-optimal actions.
We begin with the definition of $\epsilon$-Best response chain, which represents the paths of strategic improvements made through $\epsilon$-optimal responses.  In the following, $\delta_i \in \A_i$, $\bdelta \in \prod_{i} \A_i$, etc.

\begin{definition}[$\epsilon$-Best response $n$-hop chain]\label{def_epsilon_BR_chain}
 We say $\bdelta \to \bdelta'$, if there exists a player $i$ with $\delta_i \ne \delta_i'$,  $\delta_j = \delta'_j$ for all $j \ne i$ and  the pair of action profiles $\bdelta,\ \bdelta'$ satisfy the following two conditions with respect to unilateral deviations of player $i$: 
\begin{eqnarray*}
    \U_i (\bdelta') &>& \U_i (\bdelta)  \hspace{31mm} \mbox{ --- we have strict improvement  for player $i$,  and,} \\
    U_i (\bdelta') &>&  \U_i (\delta, \bdelta_{-i}) - \epsilon  \ \forall \ \delta \in \A_i,  \hspace{3mm} \mbox{ --- this move is $\epsilon$-best for  player~$i$.}
\end{eqnarray*}
A set of points $(\bdelta_1, \bdelta_2, \cdots, \bdelta_n)$ is called $\epsilon$-best response $n$-hop chain or briefly $(n,\epsilon)$-hop chain, if for every $l<n$,  we have $\bdelta_l \to \bdelta_{l+1}$. 
\end{definition}

An $(n,\epsilon)$-hop chain captures finite-step improvements. To analyse game dynamics, we require a notion of paths originating from a specific action profile. This leads us to define $(n, \epsilon)$-tails, which represent chains starting from a given action profile.

 \begin{definition}[$(n, \epsilon)$-tail] \label{defn_k_epsilon_chain}
  We say $(\bdelta_1,\cdots, \bdelta_{n})$ is a $(n,\epsilon)$-tail of $\bdelta$ if $(\bdelta, \bdelta_1,\cdots, \bdelta_{n})$ is a $(n+1, \epsilon)$-hop chain.    
 \end{definition}
 
Here, $(n,\epsilon)$-tails describe general deviation sequences. However, one would require to capture the sequential deviations starting from a set to the outside of the same.
%
To formalize this, we define external tails, that capture the deviations starting from a set~$\E$ to the outside of it, where the number of violating components increases with each step (we utilize these external chains to extend the notion of stability on the  `measure-zero'   points of violation).


 \begin{definition}[External $(n, \epsilon)$-tail] \label{defn_k_external_epsilon_chain}
 Given any product set of strategy profiles $\E = \prod_i \E_i $ and any $\bdelta \in \E$, 
     a $(n,\epsilon)$-tail $(\bdelta_1, \bdelta_2,\cdots, \bdelta_{n})$ of $\bdelta $  is called an \textit{external $(n,\epsilon)$-tail}  if  $\bdelta_i  \notin \E$ for each $i \ge 1$, which further satisfy (the number of components  outside $\E$ after $n$-unilateral moves are at least $n$):
     $$
    | \{ i : \delta_{m,i}  \notin \E_i \} | = m \mbox{ for each } m \le n, \mbox{ where } \ \bdelta_m = (\delta_{m,i})_i \mbox{ for each } m.
     $$
 \end{definition}
 
 The definition of the external tail ensures that the deviations progressively `escape' the product set under the consideration, with at least $n$ players altering their moves to outside of their respective $\{\E_i\}$ sets.
Preventing 
 such strict outward progression is crucial for stability, as it prevents an indefinite chain from a set to outside of the set.

Equipped with the above definitions, we now define the $\mu$-connected equilibrium cycle ($\mu$-EC).
This structure integrates three key properties, as  in EC of \cite{EC_Paper}: \textit{stability} against outside deviations, \textit{unrest} to permit internal instability, and \textit{minimality} to eliminate redundancy. We generalize the notion in two ways: a) we allow deviations on a measure zero set, but ensure no external chains arise from the violating points. b) we restrict the EC to be composed of connected components (this additional structure eases out the difficulties in handling  the minimality condition). 

\begin{definition}[$\mu$-Connected Equilibrium Cycle] \label{def_mu_weak_EC}
Consider a strategic form game $ \left< \N, (\A_i)_{i \in \N}, (\U_i)_{i \in \N}\right>,$ where $\N = \{1,\cdots,N\}$, for each~$i,$ $\A_i$ is a nonempty set in a metric space.  
For each $i$, consider a probability measure $\mu_i$ and a set $\E_i$ such that  $\mu_i(\E_i) >0$; also ensure   $\E_i$ is finite union of connected closed  components. Let $\mu := \prod_i \mu_i$ be the product measure. 
The $\E:= \E_1 \times \E_2 \times \cdots \times \E_N \subset \A$ is called  $\mu$ connected equilibrium cycle (in short $\mu$-EC)   if  it satisfies the following  conditions:
\begin{enumerate}
    \item \textbf{[Stability]} For  each  $i$, there exists an   $\L_{-i}$ with $\mu_{-i}(\L_{-i}) = 0$.  Then for each   
 opponent action profile  $\bdelta_{-i} \in \E_{-i} \setminus   \L_{-i} $, there exists an action $\delta_i \in \E_i$ satisfying:  
    \begin{eqnarray}
    \label{Eqn_muEC_stability}
    \U_i(\delta_i, \bdelta_{-i}) &>  & \U_i ( \tilde \delta_i, \bdelta_{-i}) \quad \text{for all } \tilde \delta_i \in \A_i \setminus \E_i. 
    \end{eqnarray}
   For $\bdelta_{-i} \in \E_{-i} \cap   \L_{-i}$ and for each $\delta_i \in \E_i$, either  \eqref{Eqn_muEC_stability}  is satisfied   or there exists no external $(n, \epsilon)$-tail, with $n \ge 2$,  from $\bdelta = (\delta_i, \bdelta_{-i})$,     for every $\epsilon >0$. 
   
     \item \textbf{[Unrest] } For any action profile   $\bdelta \in \E   $:     there exists a player $i \in \N,$ an alternate action $\delta_i' \in \E_i$ and   such that,  
    \begin{eqnarray}\label{eqn_unrest_property_EC_def}
    \U_i(\delta_i', \bdelta_{-i}) > \U_i ( \delta_i, \bdelta_{-i}) \mbox{ and }\
    \U_i(\delta_i', \bdelta_{-i})  >  \U_i ( \tilde \delta_i, \bdelta_{-i}) \ \text{for all  } \tilde \delta_i \in \A_i \setminus \E_i. \hspace{3mm}
    \end{eqnarray}

    \item \textbf{[Minimality] } No closed~$\F = \F_1 \times \F_2 \times \cdots \times \F_N \subsetneq \E$ satisfies the above two conditions.
\end{enumerate}
\end{definition}
The \textit{stability} condition ensures that the players cannot unilaterally deviate to outside the $\mu$-EC set, except on a set of measure-zero.  For the violating points,  the stability condition ensures the absence of external tails.
%
The \textit{Unrest} captures the instability, where  the players remain incentivized  to deviate within, showing (forever) instability within the set. Finally, the \textit{minimality} condition states that an $\mu$-EC is a minimal set, i.e., no strict subset of a $\mu$-EC is a $\mu$-EC.

Intuitively, a $\mu$-connected equilibrium cycle represents a closed region of action profiles where the players continuously adjust strategies within $\E$ to undercut each other but cannot escape it. The measure $\mu$  ensures ``generic" stability by excluding measure-zero pathological deviations.
Also, it is important to observe that from none of the points in $\mu$-EC, there exists a $(n,\epsilon)$-best path  (of sufficient length) to outside EC for each $\epsilon > 0$ and for any $n\geq 2$. This implies all the unrest is directed within the EC and not outwards.  Finally,  if any game has an EC $\E$ as defined in~\cite{EC_Paper} with connected components, then $\E$ will also be a $\mu$-EC, for any measure $\mu$.

\subsection{Unequal sellers equilibrium---equilibrium cycle}
\label{subsec:unequal_seller_EC}
We now move towards the most important result of the paper, the main theorem that establishes the conditions under which a $\mu$-EC exists (we mainly require~\eqref{eqn_assump_seller_eta_different}, that the sellers are of different strengths); this theorem also provides insights into the competitive dynamics between the sellers and the platform. Towards that, we begin by defining certain quantities that in turn define the $\mu$-EC for the \ospgame~defined in Section~\ref{sec:game_theoretic_modeling}. 
Consider the following definitions, defined in a backward recursive manner, starting with the quantity corresponding to seller~$(M+1)$ (the one whose strength $\eta_{M+1}$ is the $(M+1)$-th highest among all):
\begin{equation}
\begin{aligned}
\teta_{M+1} &= \eta_{M+1} , \ \ \  \tilde \eta_M = \eta_M - (\eta_M - \eta_{M+1}) \tgam^{M-1} ,\\
\teta_{M-1} &= \max\{\eta_{M-1} - (\eta_{M-1} - \tilde \eta_{M}) \tgam^{M-2},  \eta_{M-1} - (\eta_{M-1} - \tilde \eta_{M+1}) \tgam^{M-1}\} ,
\\
\teta_{k} &= \max_{ m \in \{k, k+1, \cdots, M\} } \eta_{k} - (\eta_{k} - \eta_{M+1}  ) \tgam^{m-1}, \mbox{ and finally} , \ \ 
\teta_1 \ = \  \teta_2.
\end{aligned}
\label{eqn_teta_def}
\end{equation}
The quantities $\teta_k$ represent certain threshold commissions  for each seller. For each $k\leq M$, if seller~$k$ chooses a commission greater than $\teta_k$, even if they secure the first position in the menu, there exists a commission within the range $[\eta_{M+1}, \eta_k]$ that outperforms their chosen action (see \eqref{eqn_better_at_inside_than_above} of Appendix). This helps us to define the boundary of the EC intervals defined in Theorem~\ref{thm_EC_RS}, that formally constitute the $\mu$-EC.

Define the product measure $\mu = \prod_k \mu_k$ with $\mu_k = \mu_{leb}$, the standard Lebesgue measure for all $k \le M$ and   $\mu_k = \dirac{\eta_k}$, the Dirac measure concentrated at $\eta_k$ for each $k > M$.
We now claim and prove that  the  set defined in \eqref{eqn_EC_set_defn} is the  
$\mu$-EC (proof is in~\ref{appendix_sec:equilibrium}).
\begin{theorem}[Unequal sellers, $\mu$-EC]
\label{thm_EC_RS}
For the game defined \eqref{eqn_unified_seller_utility} in Section~\ref{sec:equilibrium}, assume that the parameters satisfy~\ref{eqn_assump_seller_eta_different}. If $M < N$ and $M \geq 2$, then the set $\E$ defined as
\begin{equation}\label{eqn_EC_set_defn}
 \E := [\eta_{M+1}, \teta_{2}]^2 \times [\eta_{M+1}, \teta_{3}] \times \cdots \times [\eta_{M+1}, \teta_{M}] \times \{\eta_{M+1}\} \times \cdots \times \{\eta_{N}\} 
\end{equation}     
is a $\mu$-EC, where for any $i\leq M,$ $\teta_i$ is defined in \eqref{eqn_teta_def}.    
\end{theorem}

The $\mu$-EC $\mathcal{E}$ identified in Theorem~\ref{thm_EC_RS}  reveals a hierarchical structure where stronger sellers (with higher $\eta_i$ values) have more flexibility in setting the commissions, while the weaker ones are constrained to lower levels due to their limited strength. A key feature of this equilibrium is the endogenous emergence (i.e., not enforced by the platform) of a threshold commission level, $\eta_{M+1}$, which acts as a threat point for all top-$M$ sellers. For each seller $1 < k \leq M$, the commission oscillates between the lower bound $\eta_{M+1}$ and the upper bound $\teta_k$. No top seller will set a commission lower than $\eta_{M+1}$,  to avoid the  risk of being pushed out of the menu by the $(M+1)$-th seller. Thus, the seller ranked $(M+1)$ serves as a persistent threat.

Interestingly, $\mu$-EC established in Theorem~\ref{thm_EC_RS} exhibits a sort of resemblance  to second-price auctions: the seller with the highest $\eta$ value, i.e., seller~$1$ (with $\eta_1$), revolves its strategies around the interval $[\eta_{M+1}, \teta_2]$. This illustrates that seller~1 is not required  to operate at the marginal cost, rather offers  a commission  which is always less than the potential of the second best seller; this aspect provides  it a clear strategic advantage. 

From the platform’s perspective, this outcome is less favorable compared to the outcome  with  equal or near-equal sellers (see Theorem~\ref{thm_equal_sellers_equilibrium}).In the latter case the sellers operate at (or near) their marginal costs and the platform extracts the maximum possible commission from each of them.  
While with unequal sellers (especially when $\eta_{M+1} $ is significantly smaller) there are two troubles: i) the sellers do not settle on a fixed price, the $\mu$-EC solution suggests the unrest with EC components; and ii) the sellers may often quote commissions that are near the threat point $\eta_{M+1}$, which can imply a significant reduction in the revenue of the platform.  Basically a bigger seller competition with more competitive sellers imply a bigger advantage for the platform. 

\subsection{Numerical Results}\label{sec:numerical_results}
To study the \ospgame, we numerically generate  the well-known best response (BR) dynamics. 
As already mentioned, our game does not admit best-responses because of discontinuities, hence we resort to $\epsilon$-best response dynamics --- in each iteration one of the sellers chooses a commission  that is $\epsilon$-optimal in response to the commissions set by other sellers.  
The simulation considers a setting with $N = 5$, $M = 3$, and~$\gamma = 0.1$. The platform follows the policy defined in~\eqref{eqn_small_gam_approx_policy} to select and rank the top-$M$ items and to set the prices accordingly, in response to the commissions set by the sellers. 
\begin{figure}[ht]
\vspace{-4mm}
    \centering
    \subfloat[\centering $N = 5$, $M = 3$]{{\includegraphics[trim = {0cm 7.5cm .6cm 8.6cm}, clip, scale = 0.36]{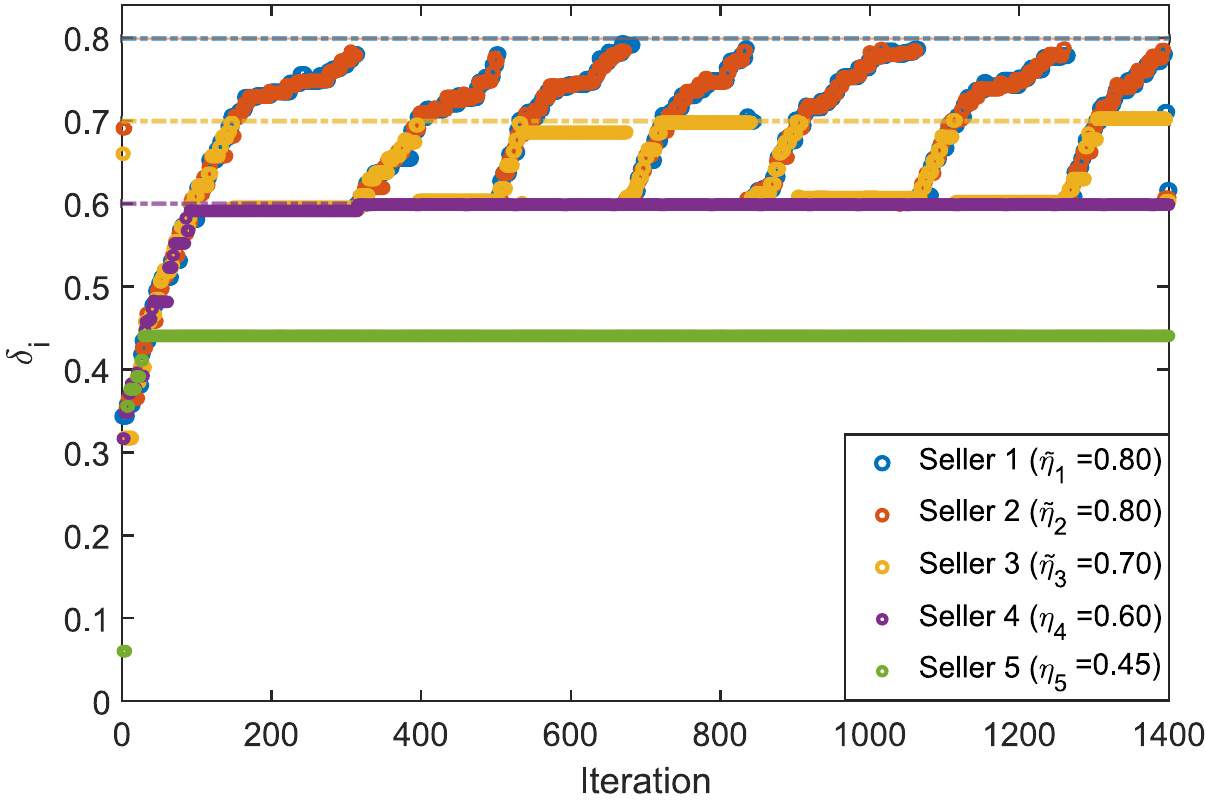}\label{fig:EC_BR_RS} }}%
    \hspace{0.2cm}
    \subfloat[\centering $N = 20$, $M = 10$ (plots for sellers 12–20 omitted for clarity)]{{ \includegraphics[trim = {0cm 7.6cm .4cm 8.6cm}, clip, scale = 0.36]{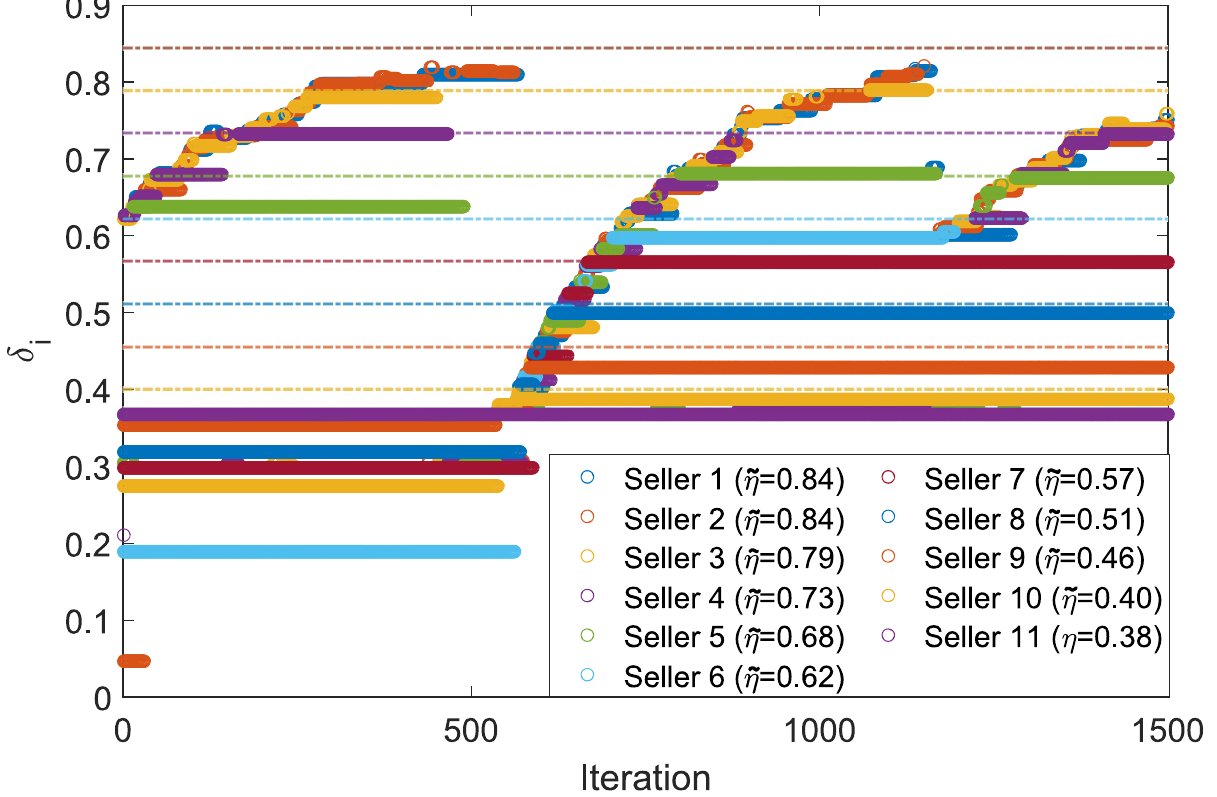}\label{fig:BR_dynamics_EC_RS_higest} }}%
    \caption{$\epsilon$-Best response (BR) dynamics of seller commissions in the \ospgame~with $\gamma = 0.1$. Horizontal dashed lines represent the theoretical thresholds \eqref{eqn_teta_def}.}%
    \label{fig:BR_dynamics_EC_RS}%
\end{figure}

\begin{figure}[ht]
\centering
\makebox[\textwidth][c]{%
  \begin{minipage}[t]{0.35\textwidth}
    \centering
    \includegraphics[width=\linewidth, trim={.3cm 6.25cm 0.3cm 5.6cm}, clip]{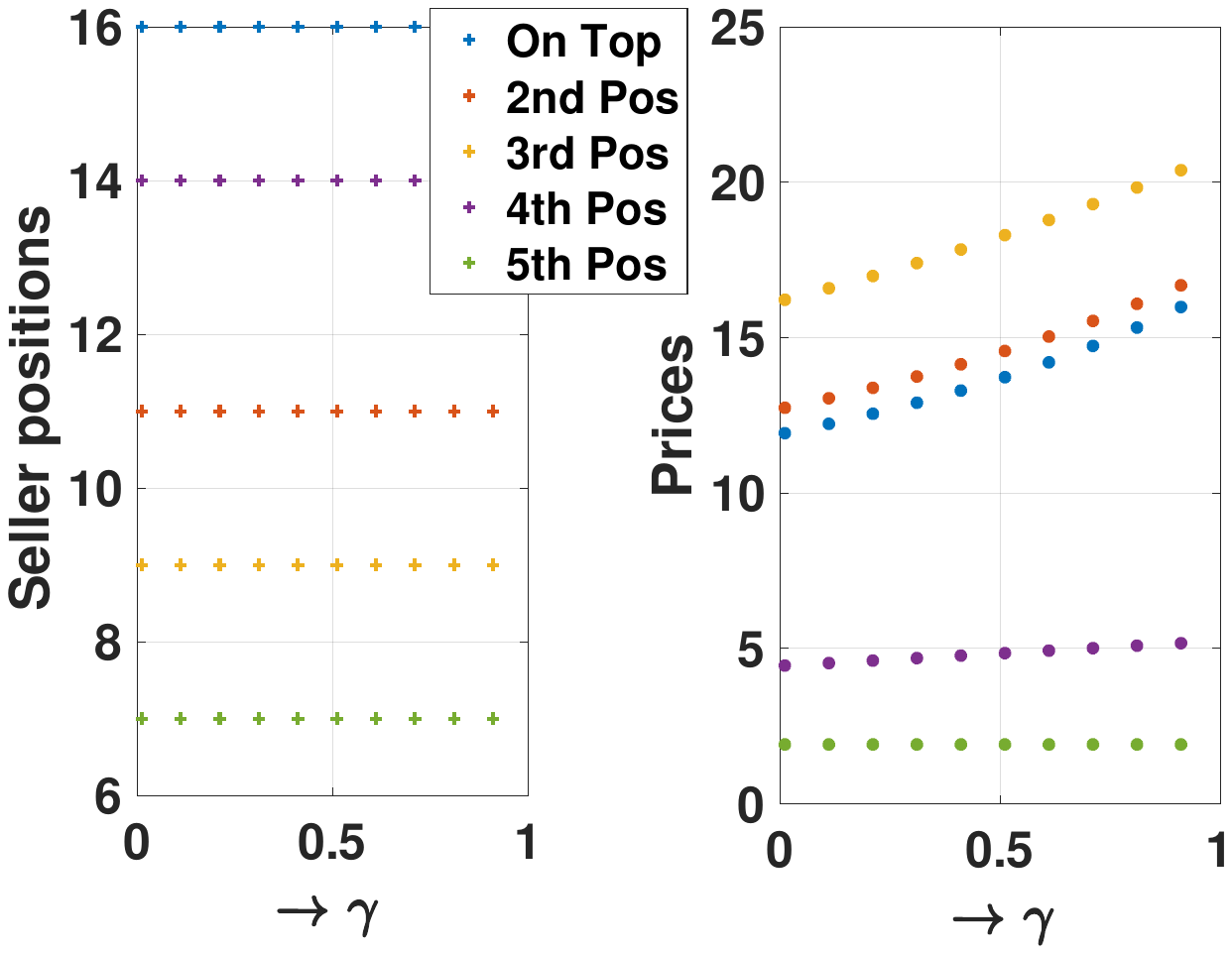}
    \captionsetup{justification=raggedright,singlelinecheck=false}
    \captionof{figure}{The optimal menu and the price vector of a random problem with $N=18$, $M=5$ for different $\gamma$.}
    \label{fig:RandomProblem}
  \end{minipage}%
  \hspace{0.02\textwidth}%
  \begin{minipage}[t]{0.65\textwidth}
    \centering
    \begin{minipage}[t]{0.48\linewidth}
      \centering
      \includegraphics[width=\linewidth, trim={0cm 5.6cm 0.4cm 3.6cm}, clip]{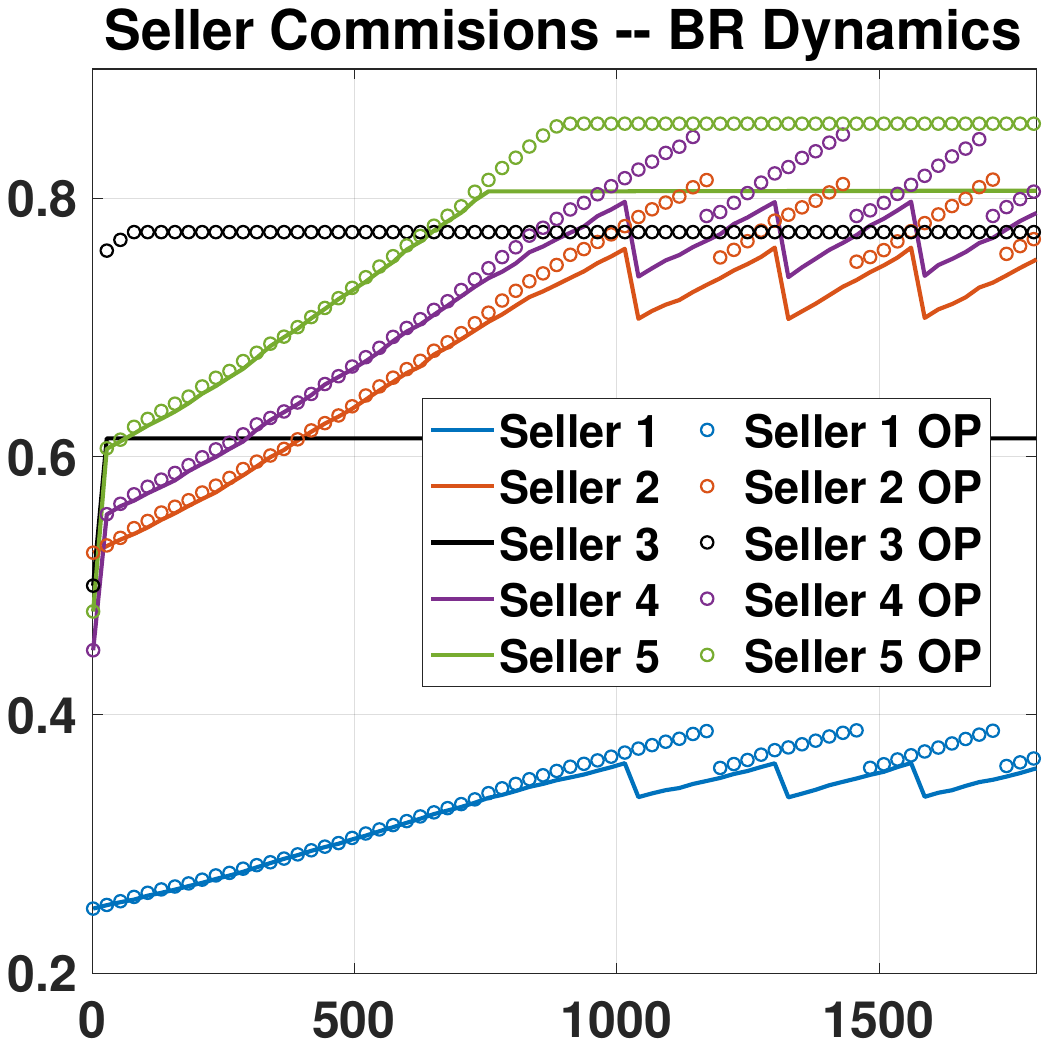}
      \captionof{figure}{OS-OP game with $\gamma=0.7$.}
      \label{fig:BR_dynamics_with_DP_7}
    \end{minipage}%
    \hfill
    \begin{minipage}[t]{0.48\linewidth}
      \centering
      \includegraphics[width=\linewidth, trim={0cm 5.6cm 0.4cm 3.6cm}, clip]{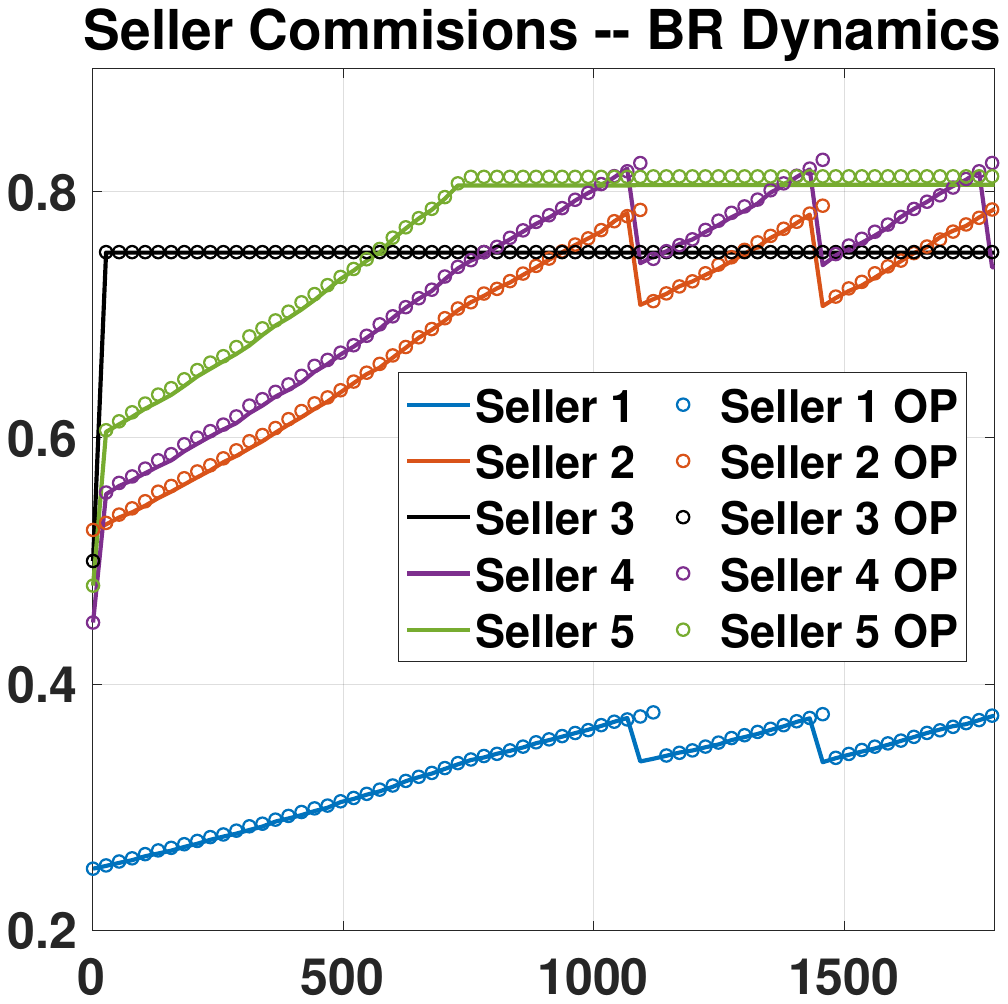}
      \captionof{figure}{OS-OP game with $\gamma=0.1$.}
      \label{fig:BR_dynamics_with_DP_1}
    \end{minipage}
    \vspace{-3mm}
    \captionsetup{justification=raggedright,singlelinecheck=false}
    \captionof*{figure}{Comparison of BR dynamics OS-OP games with $\alpha=[0.1,0.21,0.3,0.22,0.24]$,  $N=5$, $M=3$, and $c_{ma}=[1.5,0.95,0.97,0.81,1.0]$.}
  \end{minipage}%
}
\vspace{-3mm}
\end{figure}




Figure~\ref{fig:BR_dynamics_EC_RS} shows the evolution of sellers’ commission strategies over iterations. The horizontal dashed lines correspond to the analytically computed threshold values $\teta_k$ (defined in~\eqref{eqn_teta_def}) for the top \rev{$M$} positions in the menu. 
As seen in the Figure~\ref{fig:EC_BR_RS}, the top three sellers
might not secure positions in the menu in the initial iteration, but soon reach  a position where they persistently 
ensure inclusion in the top-$3$ menu.  Further, they
do not stabilize to fixed commission levels,  but instead oscillate within the interval $[\eta_4, \teta_k]$. As seen these strategic interactions have a pattern:  initially, all three sellers undercut one another until seller~3 hits the threshold $\teta_3$;  subsequently, sellers 1 and 2 continue undercutting each other until both reach $\teta_1 = \teta_2$. Once all three have reached their respective thresholds, the dynamics restarts from $\eta_4 = 0.6$, by one of them making the first move; and the cycle continues. This oscillatory pattern reflects the structure of the equilibrium cycle shown in Theorem~\ref{thm_EC_RS}.
In contrast, after some iterations (once the top~$3$ sellers reach above~$\eta_4$), sellers 4 and 5, persistently fail to grab a position in the menu. 
This aligns with the Theorem~\ref{thm_EC_RS}, which states that at equilibrium, $N-M$ sellers do not get position in the $\mu$-connected equilibrium cycle. \rev{An exactly similar pattern appears in Figure~\ref{fig:BR_dynamics_EC_RS_higest}, even when   the menu size ($M-10$) and the number of items ($N=20$) are large.}

\rev{In Figures \ref{fig:BR_dynamics_with_DP_7} and \ref{fig:BR_dynamics_with_DP_1}, we present  yet another numerical study, where the $\epsilon$-BRs (and the platform policies) are computed without using small $\gamma$-approximation \eqref{eqn_small_gam_approx_policy}; basically we solve the platform optimization problem  directly  using the DP equations \eqref{eqn_DP_equations}, while computing the BRs  of   sellers (in each iteration); we then compare the OP game   BR-dynamics with that of OS game. The parameters of the example are provided in the caption itself. Interestingly, we again have equilibrium cycles; further the OP-BR-dynamics (almost) match with that corresponding to the OS game when $\gamma$ is small (in the Figure~\ref{fig:BR_dynamics_with_DP_1} with $\gamma = 0.1$, the  OP-circle-curves almost coincide with OS straight-line-curves); more interestingly, the cyclic-structure of the two dynamics are not significantly differently even for larger $\gamma$ (in the Figure~\ref{fig:BR_dynamics_with_DP_7}  with $\gamma = 0.7$,  we again have cycles for both the cases and further   the cycle lengths are almost the same).} 

\rev{
In the final case study, we analyze the optimal policy of the platform   as a function  of $\gamma$. We consider several case studies with  different  $(N,M)$ where  different instances of    $\alpha, c_{ma}$ parameters are generated randomly (solution for one  instance is provided in Figure~\ref{fig:RandomProblem}). In all the case-studies, without exception,  we make the following striking observation --- \textit{the optimal menu in the OP game (computed by solving DP equations) does not depend upon $\gamma$ and further matches exactly with the one in the OS game (given by h-index rule, \eqref{eqn_h_index})}.  However,  the optimal price-vectors vary with $\gamma$ (see    Figure~\ref{fig:RandomProblem}). 
Once the optimal menu is  computed using the simplistic $h$-index rule,  it is  significantly easier to obtain the price-components of the optimal policy, while solving the  DP equations \eqref{eqn_DP_equations}   in backward fashion.  
}

\section{Conclusions}\label{sec:conclusions}
In this paper, we study a recommendation system where strategic sellers compete for visibility by offering commissions to a revenue-maximizing platform that curates a ranked menu of items. Our model captures two realistic platform scenarios: one where the platform designs  an optimal ranking  menu and sets optimal prices (OP game), and another where the prices are seller-prefixed  (OS game), and the platform  only  decides the ranking. By modelling these  two scenarios as two Stackelberg games, with sellers as leaders and the platform as the follower, we characterize how commission strategies influence the platform’s pricing and ranking decisions; both the games assume customer behaviour is  governed by a cascade click model.

One of the  important contributions of this work lies in formulating the platform’s problem as an MDP, enabling recursive computation of  the optimal and fair pricing and ranking policies. We derive closed-form expressions for the platform’s policy in the small $\gamma$ regime (low customer exploration regime, where customers explore only few items), which reduces computational complexity and allows us to analyse both the games under a unified framework.  We numerically illustrate that the loss of accuracy with the limit game is negligible even at moderate exploration probabilities, for example even up to $0.4$. 

\rev{The inclusion of strategic influence of sellers into the traditional cascade-click models has provided several new insights in both the games.}
Interestingly, we show that the standard pure-strategy Nash equilibria (NE)   exists only for the scenarios with equal or near equal sellers; more interestingly at the resulting NE  the sellers operate at marginal costs, exactly like in Bertrand duopoly. 
Once the sellers are of different strengths,  the outcome of the game in the space of  pure strategies, is  captured  by    a novel solution concept, namely  $\mu$-connected equilibrium cycle.  Our analysis also reveals how weaker sellers can act as strategic threats to stronger ones, pushing the latter to offer higher commissions in order to maintain visibility.

The new solution 
concept generalizes classical equilibrium notions among pure strategies and explains seller behaviour that cycles through certain stable regions of the strategy space. The proposed concept is a generalization of a recently introduced solution concept called equilibrium cycle, to provide solution for scenario with measure-zero exceptions.  
Overall, our work provides a unified and tractable framework to analyse recommendation systems with strategic sellers, bridging optimization, game theory, and customer behaviour modelling. It opens up directions for further research, including the design of fair or regulated platform policies, dynamic learning in response to customer feedback.


\bibliographystyle{elsarticle-num-names}
\bibliography{references}
\appendix
\section{Proofs related to Section~\ref{sec_MDP}} \label{Appendix_for_sec_MDP}
\begin{proof}[Proof of Lemma~\ref{lem_equivalence_between_random_deterministic_policy}]
Let $\mu = (d_1, d_2, \dots, d_M)$ be a deterministic policy. At each step $t$, the decision rule $d_t(\lx)$ specifies an action $(a_t, p_t)$ based on the current state $\lx$. Starting from the initial state $\bar \lx = \I$, the policy $\mu$ generates a sequence of actions $(a_1, p_1), (a_2, p_2), \dots, (a_M, p_M)$, where, $(a_t, p_t) = d_t(\I - \{a_1, a_2, \dots, a_{t-1}\})$. Observe that at any $t$, the transitions are only to an absorbing state $\Delta$ or to state $\lx' = \I-\{a_1, \cdots, a_{t}\}$.
Define $\bar \z^\mu$ as the sequence of actions generated by $\mu$:
$$
\bar \z^\mu = \big((a_1, p_1), (a_2, p_2), \dots, (a_M, p_M)\big).
$$
By construction, the expected reward under policy $\mu$ is $J(\mu, \bar \lx)$, which is equivalent to the revenue $R(\bar \z^\mu, \bar \lx)$ generated by the sequence $\bar \z^\mu$. This is because the policy $\mu$ and the sequence~$\bar \z^\mu$ induce the same customer behaviour and revenue outcomes. Thus, we have:
$$
J(\mu, \bar \lx) = R(\bar \z^\mu, \bar \lx).
$$

\ignore{Say $\mu$ is a Markov deterministic policy. The transitions are also deterministic.
At $t =1,$ we have $\bar \lx = \I,$ and let $(a_1, p_1) := d_1(\I)$. For each $i = 2, \cdots, N$, say $ (a_i, p_i):= d_i(\I-\{a_1, \cdots, a_{i-1}\})$.  Observe that at any $t$, the transitions are only to an absorbing state $\Delta$ or to state $\lx' = \I-\{a_1, \cdots, a_{t}\}$. Thus, consider $\bar \z^\mu$ such that
\begin{equation*}
 (a_1, p_1) = \bar \z_1^\mu,
 (a_2, p_2) = \bar \z_2^\mu, \cdots,
 (a_M, p_M) = \bar \z_M^\mu.
\end{equation*}
That is, there exists $\bar \z^\mu$ such that $J(\mu, \bar \lx)  = R(\bar \z_\mu, \bar \lx).$ This proves part~1.
}

Next, for part~(ii), consider any sequence $\bar \z = \big((a_1, p_1), (a_2, p_2), \dots, (a_M, p_M)\big)$ --- this is given by a fixed menu $\pi = (a_1, \cdots, a_M)$ and the price vector $\p = (p_1, \cdots, p_M)$. We construct a deterministic policy $\mu_{\bar \z} = (d_1, d_2, \dots, d_M)$ as follows:
For each step $t$, define the decision rule $d_t$ such that
  $$
  d_t(\lx) = \begin{cases}
    (a_t, p_t) & \text{if } \lx = \I - \{a_1, a_2, \dots, a_{t-1}\},\\
    (a_1, 0) & \text{otherwise (if $\lx = \Delta$)}.
  \end{cases}
  $$

This policy $\mu_{\bar \z}$ ensures that the sequence of actions taken under $\mu_{\bar \z}$ matches $\bar \z$. Consequently, the expected reward under $\mu_{\bar \z}$, $J(\mu_{\bar \z}, \bar \lx)$, is equal to the revenue $R(\bar \z, \bar \lx)$ generated by $\bar \z$. Thus, we have:
$$
R(\bar \z, \bar \lx) = J(\mu_{\bar \z}, \bar \lx).
$$

\ignore{Consider any $\bar \z $. For any $\lx \subseteq \I$, choose deterministic policy $\mu_{\bar \z}$ such that
$$
d_i( \lx) = \begin{cases}
    \bar \z_i & \text{if }  \lx \neq \Delta,\\
    (a_1, 0) & \text{else}.
\end{cases}
$$
That is, there exists a policy $\mu_{\bar \z} = (d_1, d_2, \cdots, d_N)$ such that $R(\bar \z, \bar \lx) = J(\mu_{\bar \z}, \bar \lx).$ This proves part~2.}

From parts (i) and (ii), we have established a one-to-one correspondence between the deterministic policies $\mu$ and the fixed menu-price vectors $\bar \z$, immediately implying equality $a$ in part~(iii); supremum is replaced by maximum as the domain $\{\pi\} \times   [0, p^\text{max}]^N$ is compact (here $\{\pi\}$ is the set of permutations or all menus of size $M$ from $N$ items). 
%
%
\end{proof}

\section{Proofs related to Section~\ref{sec_small_gamma_game}} \label{Appendix_sec_small_gamma_game}
\label{sec_Appendix_sec_four}
{
\begin{proof}[Proof of Theorem~\ref{thm_small_gamma_optimality}]
We use backward induction to derive this proof. We will prove that at any $t = M - k,$ there exists a $\bar \gamma_k > 0,$ such that for all $\gamma < \bar \gamma_k, $ the optimal permutation is given by \eqref{eqn_opt_perm_small_gamma}, when the initial condition at $t=1$ is  $\I$ (it is clear that under the given hypothesis of theorem, we have unique such order). Towards proving this, it suffices to prove  the following for at any $t$ and $\lx$. At any $t=M-k$,  there exists a $\bar \gamma_k >0$ such that for all  $\gamma < \bar \gamma_k$  the following unique optimizer (pair) solves the DP equation (recall $a_1(\lx)$ is the unique seller item with highest $f^*$-index among $\lx$, and $p_t^*(\lx)$ is   unique under Assumption~\ref{assumption_beta_fun}):
\begin{eqnarray}    
\label{Eqn_induction_step}
\pi_t^*(\lx) = a_1(\lx) \mbox{ and }
\{p^*_t(\lx)\} = \mbox{\Arg} \max_p \left \{ \delta_{a_1(\lx)}\beta_{a_1(\lx) } (p) (p -  \gamma v_{t+1} (\lx \setminus \{a_1(\lx) \} ) \right \}.
\end{eqnarray}

\textit{Base step at $t = M$:} 
Equation \eqref{Eqn_induction_step} is true 
trivially,   by strict monotonicity assumed in the hypothesis of this theorem: basically we have   $\pi^*_M(\lx) = (a_1(\lx))$ for all $\lx$ and thus have unique optimizer $\z_M^*(\lx) =  (a_1(\lx), p^*(a_1(\lx)) )$, where   $p^*(a) $ is the unique optimizer. 


\textit{Inductive hypothesis at $t = M - (k-1)$}: there exists a $\bar \gamma_{k-1} > 0,$ such that for all $\gamma < \bar \gamma_{k-1}, $ 
the optimal choice  is given by  
given by \eqref{Eqn_induction_step} for any $\lx$.

\textit{Inductive step at $t = M - k$}: At $t = M - k$ consider any $\lx$; using \eqref{eqn_DP_equations}, for any $a \in \lx,$ define the partial Q-factors as below (observe these are not standard Q-factors as we only consider $a$ and not the second component of action $p$ --- in fact, we actually optimize over $p$),
{\small
$$
Q^*_{M-k}(a,p) := \max_{p \in [0, p^{\text{max}}]} \big( \delta_a \beta_a(p) p + \gamma (1 - \beta_a(p)) v_{M-k+1}(\lx \setminus \{a\}) \big);
$$
}%
note the above maximization problem has unique $p^*$ as optimizer by Assumption~\ref{assumption_beta_fun}. 
Observe that for any $a \in (\lx\setminus\{a_1(\lx)\}),$
{\small
\begin{equation}\label{eqn_diff_small_gamma}
\max_p Q^*_{M-k}(a_1(\lx),p) - \max_p Q^*_{M-k}(a,p) = C(\gamma) + C_1.    
\end{equation}
}
Here, $C(\gamma)$ is function of $\gamma$ and using hypothesis that for all $a \in (\I\setminus\{a_1(\I)\})$, $\fstar(a_1(\I) ) > \fstar(a)$ we have  $C_1 > 0$. At $\gamma = 0,$ the RHS of~\eqref{eqn_diff_small_gamma} is strictly positive. Thus, for each $a \in \lx\setminus\{a_1(\lx)\}$, there exists a $\bar \gamma_k^{a} > 0,$ such that for all $\gamma < \bar  \gamma_k^{a},$ the RHS of \eqref{eqn_diff_small_gamma} is positive. Thus, by choosing $\bar \gamma_k = \min_{a \in \lx}  \bar \gamma_k^{a},$ the RHS of \eqref{eqn_diff_small_gamma} is positive for any $a \in (\lx\setminus\{a_1(\lx))\}$ and $\gamma <  \gamma_k$ (choose this new $\bar \gamma_k $ lesser than  or equal to each $\bar \gamma_{k'}$, where   $t'=M-k' > t$.)  Thus,   for $\gamma <  \bar \gamma_k$, the optimal permutation is as given in~\eqref{Eqn_induction_step} even at $t = M-k$.

Finally, choose $\bar \gamma = \min_k \gamma_k = \bar \gamma_1,$ then for all $\gamma < \bar \gamma$, the optimal permutation is as given in~\eqref{Eqn_induction_step} and hence as in \eqref{eqn_opt_perm_small_gamma} when at $t=1$ the initial state is $\I$. This completes the proof. 
\end{proof}
} 

\section{Proofs related to Section~\ref{sec:game_theoretic_modeling}} \label{appendix_sec:game_theoretic_modeling}
\begin{proof}[Proof of Lemma~\ref{lem_OS_OSP_similarity}]
From expressions \eqref{eqn_item_prices_OS_game}  \eqref{eqn_small_gam_approx_policy} and \eqref{eqn_opt_price_and_beta_approx}, the optimal price of any item~$a$ is given by $\nicefrac{1}{\alpha_a}$. Substituting this into the expressions for the h-index and $\fstar$-index, we get:
$$
h(a, p^*_a) = \frac{\delta^*_a}{ \alpha_a (1 + e(1 - \gamma))},  \quad \text{ and }
f^*(a) = \frac{\delta^*_a}{ \alpha_a e}.
$$
Since $e$ is a constant and $\gamma$ is fixed for any game instance,  both indices rank sellers according to the same quantity~$\nicefrac{\delta^*_a}{\alpha_a}$. Thus, the seller rankings---and consequently their utilities---are the same in both the OS game and the \ospgame.
\end{proof}

\section{Proofs related to Section~\ref{sec:equilibrium}} \label{appendix_sec:equilibrium}
\begin{proof}[Proof of Theorem~\ref{thm_equal_sellers_equilibrium}]
\textbf{Part~(i):} Observe that at $\bdelta_*$, top $L$ players share the top~$M$ positions in an equally likely manner. Thus, each player $k \le L$ receives utility (see \eqref{eqn_unified_seller_utility}):
\begin{eqnarray} \label{eqn_all_players_same_delta}
    \U_k (\bdelta_*) = \frac{\eta_c - \delta_{*,k}}{e \alpha_k}  \left( \frac{1 - \tgam^{M}}{L(1-\tgam)}   \right) = 0  \mbox{ as } \delta_{*,k} = \eta_c \mbox{ for all } k \le L.
\end{eqnarray}

On the other hand, any player with $k > L$ derives zero utility at respective $\delta_k$ as they do not get a position in the menu.  
If they 
deviate to a $\delta_k'$ (against $\bdelta_{*,-k}$) that gets them a position in the menu, from~\eqref{eqn_unified_seller_utility}, they derive strictly negative utility. 

We now consider all possible unilateral deviations by a player $k \le L$.
Suppose player $k$ increases its commission from $\eta_c$ to $\eta_c + \varrho$ for any  $\varrho > 0$, then it gets the first position in the menu exclusively,  however, receives negative utility:
\begin{eqnarray*}
\U_k (\eta_c + \varrho, \bdelta_{*,-k} ) = 
    \frac{ - \varrho}{e \alpha_k}  < 0. 
\end{eqnarray*}
If player $k$ unilaterally decreases its commission to a  $\delta_k' < \eta_c$, then using~\eqref{eqn_unified_seller_utility}, this action pushes player~$k$ out of the top~$M$ slots. As a result, it receives zero utility. Thus, $\bdelta_*$ is a Nash equilibrium (see~\cite{narahari2014game}).

\textbf{Part~(ii):} Consider any player $k < (M+1)$, then using \eqref{eqn_unified_seller_utility},  $\U_k (\bdelta_*) = (\nicefrac{(\eta_k-\eta_{k+1})}{e \alpha_k}) \tgam^{k-1}$. When it deviates unilaterally to an alternate choice $\delta_k' > \eta_{M+1}$,  it derives another position $k'\in \{1, \cdots, M\}$ depending upon $\delta_k'$, however its  utility does not improve by more than $\epsilon$ as,
\begin{eqnarray*}
    \mbox{ using \eqref{eqn_unified_seller_utility}, we have } \ \ \U_k (\delta_k',  \bdelta_{*,-k}) = \frac{\eta_k-\delta_k'}{e \alpha_k} \tgam^{k'-1}   < \epsilon ,
\end{eqnarray*}
 under the given hypothesis. With $\delta_k' \le \eta_{M+1}$: (i) it either does not get a position in the menu and derives zero if $\delta_k' < \delta_l$ for some  $l \ge (M+1)$; or (ii) 
 possibly shares the $M$-th position with   $(M+1)$-th player but does not derive more than $\epsilon$.

This completes the proof using similar logic as in part (i) for players  $k \ge (M+1)$.
\end{proof}

\begin{proof}[{\bf Proof of Theorem~\ref{thm_EC_RS}}]
Before proceeding,   we prove an important condition in \eqref{Eqn_a_gt_teta_i_inferior}, given below,  that is going to  be used throughout the proof. 

By definition in \eqref{eqn_teta_def}, for any player $k > 1$, any $\delta > \teta_k$, we have:
\begin{eqnarray}
    \eta_k - \delta <    (\eta_{k} - \eta_{M+1}  ) \tgam^{m-1}   \mbox{ for each } m \in \{ k, \cdots, M\}.  \label{Eqn_a_big_teta_not_good}
\end{eqnarray}
Thus from \eqref{eqn_unified_seller_utility}, for seller $k$,
  choosing a $\delta > \teta_k$ and securing even the first position is inferior to choosing   one among $\eta_{M+1}$ (or $\eta_{M+1}+\varrho$ for $\varrho$ sufficiently small, by strict inequality in~\eqref{Eqn_a_big_teta_not_good})  and securing the $m$-th position (alone). \textit{This also implies that, against any action profile of the opponent players $\bdelta_{-k}$, any $\delta > \teta_k$ is an inferior choice for seller $k$ (even  if it secures  the first position in the menu) --- 
   it derives better, for example, by choosing the following and securing the $M$-th position in the menu,}
\begin{eqnarray}\label{eqn_better_at_inside_than_above}
       \U_k(\eta_{M+1} +\varrho, \bdelta_{-k} )  > \U_k (\delta, \bdelta_{-k} ) \mbox{ for some small enough } \varrho < \min_{j \ne k, j \le M} \delta_j. 
\end{eqnarray}
In the above,  $\varrho >0 $ is small enough to satisfy \eqref{Eqn_a_big_teta_not_good} with $m = M$ and even after replacing $\eta_{M+1}$ with  $\eta_{M+1} +\varrho$.

Further, for player~1,  there is no real incentive to choose any $\delta > \teta_2,$ 
as a $\delta' =  \teta_2 $ would  be strictly better (unless $\delta_2 = \teta_2$). Recall, $\teta_1 = \teta_2.$ 
Thus in all, for any $k\le M$   and any $\bdelta_{-k} $, 
\begin{eqnarray}
\mbox{ for any $\delta > \teta_k $,  \  there exists an $\delta' \in [\eta_{M+1}, \teta_k]$ \ such that, }
  \U_k (\delta, \bdelta_{-k} ) < U_k (\delta', \bdelta_{-k}),  \hspace{6mm}\label{Eqn_a_gt_teta_i_inferior}
\end{eqnarray}
\textit{unless $k=1$ and $\delta_2 = \teta_2.$}

In the proofs given below we use $i,k,l$ as player indices. 
\medskip

\subsection*{  Proving the stability property \eqref{Eqn_muEC_stability}} For  any $i > M$, we have $\E_i = \{\eta_i\}$ and the agent derives zero utility at the singleton option, which is the maximum that it  can derive, see \eqref{eqn_unified_seller_utility}. 
Thus,
it suffices to prove the condition 1 for players $i \le M$ and
consider one such player $i$ and then any $\bdelta_{-i} \in \E_{-i}$.

Since $\delta_{M+1} = \eta_{M+1},$ for any $\tilde \delta_i < \eta_{M+1}$, player~$i$ would not get any of the first $M$  positions and hence using \eqref{eqn_unified_seller_utility}, we have:
\begin{equation}\label{eqn_dominating_below} \U_i(\tilde \delta_i, \bdelta_{-i}) = 0 \ \mbox{ for any } \tilde \delta_i < \eta_{M+1}. 
\end{equation}
Now for $i > 1$  using \eqref{Eqn_a_gt_teta_i_inferior}, we have (choosing $\varrho$ as in \eqref{eqn_better_at_inside_than_above}),
\begin{equation}\label{eqn_dominating_above_for_general}
\U_i( \delta_i, \bdelta_{-i}) >  \U_i(\tilde \delta_i, \bdelta_{-i})  ,  \mbox{ with } \delta_i = \eta_{M+1}+\varrho \mbox{ and any }  \tilde \delta_i > \teta_i, 
\end{equation}
proving the first property  for any player $i > 1$ with $ \delta_i $,  also using \eqref{eqn_dominating_below} --- because from \eqref{eqn_unified_seller_utility}, 
$$
\U_i(\delta_i, \bdelta_{-i}) >  \frac{(\tilde \eta_i - \eta_{M+1} - \varrho) \tgam^{M-1}}{e \alpha_i} > 0.
$$


For player~$1$,  define $\L_1 = \{\teta_{2}\} \times \prod_{j = 3}^N \E_j$, note  $\mu_{-1} (\L_1) =0$. Now, for any   $\bdelta_{-1} \in \E_{-1} \setminus \L_1$ (note that $\max_{j \ge 2} \delta_j < \teta_{2}$), choose $\delta_1 = \teta_{2}$, and thus we have:
\begin{equation}
\label{eqn_P_1_domination_above}
\U_1(\delta_1, \bdelta_{-1}) = \frac{ \left( \eta_1 -  \delta_1 \right)  }{e \alpha_1}  > \frac{ \left( \eta_1 -  \tilde \delta_1 \right)  }{e \alpha_1} = \U_1(\tilde \delta_1, \bdelta_{-1})  > 0 ,  \quad  \forall \ \tilde \delta_1 \in (\delta_1, 1].  
\end{equation}
Thus we could establish the first (stability) property for all $i$ and all $\bdelta_{-i}$, except for $i=1$ (player 1) and for $\bdelta_{-1}$ with $\delta_2= \teta_2$. 
 In the below, we show that there exists no external tails from such $\bdelta \in \E$ which completes the proof of stability property. 



%
\medskip
\noindent
{\bf External tails:}
Consider any $\bdelta \in \E$ such that $\bdelta_{-1} \in \L_1$.
Now, player~1 can only deviate to one of the following points (no other deviation is possible), that is to the potential set of improvements $ {\cal I}_{\epsilon,1}$,  by choosing appropriate an $\varrho > 0$  in `$\epsilon$'-optimal way:
$$
\delta_1' \in {\cal I}_{\epsilon,1} :=   \bigg \{  \delta_j + \varrho:  \delta_j   < \teta_2  \bigg\} \cup \{\eta_{M+1} + \varrho\} \cup \{ \teta_2 + \varrho\}. 
$$
Observe  here that one can choose $\varrho $ sufficiently small such that $(\delta_j + \varrho) \in  [\eta_{M+1}, \teta_2] $ for each   $\delta_j  < \teta_2$.
Observe also that the above are the only types of improvement sets\footnote{For bigger values of $\epsilon$, one might find an external $(k,\epsilon)$ chain, but not for smaller $\epsilon$, depending upon $\bdelta$.} for every $\epsilon >0$ lesser than some small enough $\bar \epsilon > 0$. 
We reiterate here that no other deviation is possible to improve in $\epsilon$-optimal way (for every $\epsilon>0$).

If for some $\bdelta_{-1} \in \L_1$,  among the moves given in $\I_{\epsilon, 1}$ there exists a $\delta_1' \neq \teta_2 + \varrho$ which provides the best improved utility, then \eqref{Eqn_muEC_stability} is satisfied with that $\delta_1'$ against that  $\bdelta_{-1}$; in fact this is true for all $\epsilon$ sufficiently small. 

Now consider a $\bdelta_{-1} \in \L_1$ for which $\delta_1' = \teta_2 + \varrho$ provides the best improved utility among all possible choices in $\I_{\epsilon, 1}$. 
In this case, $\bdelta \to \bdelta_1 := ( \teta_2 + \varrho, \delta_2, \cdots, \delta_N)$, where $\bdelta_1 \notin \E$; also the number of external components  $|\{ i: \delta_{1, i} \notin \E_i \}| = 1$.  In response, for example, player~2 has strict incentive to play an action $\delta_2' = \max_{k >2} \delta_k + \varrho,$ for sufficiently small $\varrho > 0.$ That is, we have $\bdelta_1 \to \bdelta_2 := ( \teta_2 + \varrho, \delta_2', \delta_3, \cdots, \delta_N)$. Observe that $\bdelta_2 \notin \E$, however, the number of external components have not changed after the second move, i.e., $|\{ i: \delta_{2, i} \notin \E_i \}| = 1$ (in fact after such a move by player~$2$ player~$1$ has strict incentive to move back to interior of $\E_1$). Strict improvement by players other than $2$ is also possible, but even such improvements lead to a point with only $\delta_1'$ being outside its corresponding EC component, $\E_1$.
This 
contradicts the requirement of external $(n,\epsilon)$, with $n \ge 2$,  chain as given in Definition~\ref{def_mu_weak_EC} (one again this is true for all~$\epsilon$ lesser than a threshold, as for such~$\epsilon$, the $\epsilon$-best component $\delta_2' \in \E_2$). 

In all, one can have
first unilateral improvement move to the external of $\E$ (when player~$1$ responds to some points in~$\L_1$), but not the subsequent moves.

\subsection* {Proving the unrest property~\eqref{eqn_unrest_property_EC_def}:} 
Consider any~$\bdelta \in  \E.$ Choose any $k \in \Argmax_j \delta_j$ and then choose any $l \in \Argmax_{j\neq k} \delta_j.$  We now proceed case-wise as given below:
\begin{enumerate}[$\bullet$]
    \item If $\delta_k = \delta_l$ (the first maximum equals the second), 
    and further say  $\delta_k < \teta_{2}$. Define $b := |\Argmax_j \delta_j |$, note $b \ge 2$.  
    
    Now player~$1$ can appropriately choose $\delta_1'  \in (\delta_k, \teta_{2})$  to improve its utility as:
    $$
    \U_1(\delta_1, \bdelta_{-1})
    \stackrel{a}{\le}  \frac{ \left( \eta_1 -  \delta_1 \right)  }{e \alpha_1} \frac{\sum_{0 \leq j < \b} {\tgam}^{j} \indic{j \le M} }{b}
    < \frac{ \left( \eta_1 -  \delta_1' \right)  }{e \alpha_1} = \U_1(\delta_1', \bdelta_{-1}).
    $$
 In the above, if $\delta_1 = \delta_k$, we have  equality in `$a$'; otherwise, we have a strict inequality in `$a$', because  for any $n > b$, (which represents the position of seller 1 in the menu with $\bdelta$), 
$$
\tgam^{n} \le  \frac{1 + \tgam + \tgam^2 + \cdots + \tgam^{ b}}{ b} < 1, \  \mbox{ as } b \ge  2.
$$
Also, observe  $\U_1(\delta_1', \bdelta_{-1}) > \U_1(\teta_{2}, \bdelta_{-1})$ and further using~\eqref{eqn_dominating_below}, the remaining part of the second condition \eqref{eqn_unrest_property_EC_def} is satisfied for this sub-case.

\item If $\delta_k = \delta_l = \teta_2$, it implies $\delta_1=\delta_2 = \teta_2$;   recall by EC definition, $\delta_i \le \teta_3$ for all~$i > 2$.  Now, player~$2$ improves  by choosing $\delta_2' = \teta_3 + \varrho'$   for some sufficiently small~$\varrho'>0$ (irrespective of $\delta_3$, etc),  see \eqref{eqn_unified_seller_utility}:
\begin{eqnarray*}
    \U_2 (\bdelta) = \frac{\eta_2 -  \teta_2}{e \alpha_2} \left(\frac{1 + \tgam}{2}\right) < \frac{\eta_2 -  \teta_2}{e \alpha_2} \stackrel{\mbox{ by \eqref{eqn_teta_def} }}{\le} \frac{\eta_2 -  \teta_3 - \varrho' }{e \alpha_2} \tgam =  
\U_2(\teta_3 + \varrho', \bdelta_{-2}).
\end{eqnarray*}
The second part of \eqref{eqn_unrest_property_EC_def} follows
 using \eqref{eqn_dominating_below}-\eqref{eqn_dominating_above_for_general}, if required  by replacing $\delta_2'$ with $\eta_{M+1}+\varrho$ with sufficiently small $\varrho>0$, as in \eqref{eqn_better_at_inside_than_above}; the replacement is required if  $\U_2(\teta_3 + \varrho', \bdelta_{-2}) < \U_2(\eta_{M+1} + \varrho, \bdelta_{-2})$.

\item If $\delta_k > \delta_l$ then player~$k$ can  choose any $\delta_k' \in (\delta_l, \delta_k)$ and strictly improve:
    $$
    \U_k(\delta_k, \bdelta_{-k}) = 
    \frac{ \left( \eta_k -  \delta_k \right)  }{e \alpha_k} 
    < 
    \frac{ \left( \eta_k -  \delta_k' \right)  }{e \alpha_k} 
    = \U_k (\delta_k', \bdelta_{-k}).
    $$
   The second part of~\eqref{eqn_unrest_property_EC_def} follows as in the previous sub-case. 
\end{enumerate}

This finishes the proof of the second unrest property~\eqref{eqn_unrest_property_EC_def}.

\subsection*{Proving the minimality property}

In this proof, we establish the minimality of the EC set using a contradiction-based argument. For the purpose of contradiction, assume a closed set $\E^o = \E_1^o \times \cdots \times \E_N^o$ with each $\E_i^o \subset \E_i$, that satisfies the first two properties of EC. 

By hypothesis~\eqref{eqn_EC_set_defn}, observe $\E_i^o = \{\eta_i\}= \E_i$ for each $i > M$, and  for $i \le M$, we have:    
\begin{equation}\label{eqn_singleton_points}
\E_i^o = \E_i^{c} \cup  \{b_{i,1}, b_{i,2} \cdots, b_{i, n_i}\} \mbox{ with }  \E_i^o \subset \E_i,    
\end{equation}
for some $n_i < \infty$, and  some $\E_i^{c}$
 a finite union of closed intervals.
Let 
\begin{equation} \label{eqn_uu_ou_defn}
\uu_i := \min \E_i^c \text{ and } \ou_i := \max \E_i^c \mbox{ for each } i.   
\end{equation} 

We obtain the proof in two steps: (i)
we first show that all players $i\le M$ share the same left endpoint, i.e.,  $\uu_i =\uu$ and that the common $\uu = \eta_{M+1}$;
(ii) we then show that $\ou_i \ge \teta_k$ for each $k \ge i$ in the backward recursive manner, i.e., starting from $k=M+1$; also show that each $\E_i^c$ is a single connected component.\textit{ For simpler notation, henceforth for all $i $ implies for all $i \leq M.$}

\subsubsection*{Step 1: The  left endpoint~$\uu_i =  \eta_{M+1}$ for each $i$}

We first show that $\uu_i = \uu$. 
For the purpose of contradiction, assume there exist at least two players $i, j$ such that $\uu_i  < \uu_j$.
Define the following:
\begin{eqnarray}\label{eqn_def_Z_left_end_points}
   \Z := \Argmax_l {\underline u}_l, \ \  \Z_1 := \Argmin_{i \in \Z} \bar u_i \ \mbox{ and  pick any }  i^* \in \Z_1.
\end{eqnarray}
Then, by our assumption and construction, we have:
$$
\underline{u}_{i^*} = \max_l \underline{u}_l   >  \max_{l \notin \Z} \underline{u}_l := \underline{u}_{s}^*, \mbox{ the second maximum of lower bounds.}
$$
Define the following set, for some small enough $ \varrho > 0$ that ensures $\Delta_{-i^*} \subset  \E_{-i^*}^o$
$$
\Delta_{-i^*} := \left \{ \bdelta_{-i^*}= (\delta_l)_{l \ne i^*}: \delta_{l} \in (\ou_l - \varrho , \ou_l)   \mbox{ for all } l \in \Z - \{i^*\},  \delta_l \in (\uu_l, \uu_l + \varrho) \mbox{ for all } l \notin \Z \right \};
$$
this is possible by definition  of $\{\uu_i\}$ and $\{\ou_i\}$ in \eqref{eqn_uu_ou_defn}  and  note $\mu_{- i^*} (\Delta_{-i^*}) > 0.$

 \underline{Sub-case 1: assume $|\Z_1| = 1$} in \eqref{eqn_def_Z_left_end_points}, 
then  for any $\bdelta_{- i^*} \in \Delta_{-i^*}^o$  and $\delta_i \in [\underline{u}_{i^*},  {\bar u}_{i^*} ] ,$
\begin{eqnarray}
\label{eqn_left_end_points_contrad_arg} 
\delta_l \notin  \left ( {\underline u}_s^* + \varrho, \  \min_{l \in \Z, l \ne i^*}  \bar u_l - \varrho \right ) \mbox{ for all } l \ne i^*  \mbox{ and } [\underline{u}_{i^*},  {\bar u}_{i^*} ]  \subsetneq   \left ( {\underline u}_s^* + \varrho,   \ \min_{l \in \Z, l \ne i^*}  \bar u_l - \varrho \right )  \hspace{4mm}
\end{eqnarray}
and    $\underline{u}_{i^*} >  {\underline u}_s^* + \varrho$. Thus,  for any $\delta \in \E_{i^*}^c $, in fact, for any $\delta \in [\underline{u}_{i^*},  {\bar u}_{i^*} ]$,  we have the following which contradicts property 1:
\begin{equation*}
\U_{i^*}(\delta, \bdelta_{-i^*})  < \U_{i^*}(  {\underline u}_s^* + \varrho +\varrho', \bdelta_{-i^*}) \mbox{ for all } \bdelta_{-i^*} \in \Delta_{-i^*},
\end{equation*}
for some small enough $\varrho' >0$  as:
\begin{enumerate}[$\bullet$]
\item player~$i^*$ can get the same position in the menu (against all such $\bdelta_{-i^*}$) by paying a smaller commission $  {\underline u}_s^* + \varrho + \varrho'$ compared to any other $\delta \in [\underline{u}_{i^*}, \  {\bar u}_{i^*} ]$ and this is primarily because $ \underline{u}_{i^*} = \max_l \underline{u}_l   >  \max_{l \ne i^*} \underline{u}_l = {\underline u}_s^* $; 
    \item 
the singleton points $\{b_{i^*, 1}, \cdots, b_{i^*, n_{i^*}}\}$ in \eqref{eqn_singleton_points} can be shown to be inferior too, by reducing~$\varrho$ further small if required and ensuring   $b_{i^*, k} \notin ({\underline u}_s^*, {\underline u}_s^*+\varrho)$ for all $k \le n_{i^*}$; further, if any singleton $b_{i^*,k}$ improves by deriving a better position in the menu for player~$i^*$, against some $\bdelta_{-i^*} \in \Delta_{-i^*},$ then one can choose a point nearby and not in $\E_i^o$ which performs strictly better (as singletons are isolated); and also, because $\teta_{i^*}$ is still inferior by \eqref{eqn_teta_def}, even if some $\b_{i^*, k}= \teta_{i^*}$;   if some $\b_{i^*, k}= \eta_{M+1}$,
player $i^*$ can strictly improve by moving to $\eta_{M+1} + \varrho''$ (for some appropriate $\varrho'' >0$) and securing the last position in the menu solely (at $\b_{i^*, k}$ it would share position with player $M+1$).
\end{enumerate}

\underline{Sub-case 2, when $|\Z_1| > 1$:} If the best position in menu derived by $i^*$ by choosing a $\delta_{i^*}$ near $\ou_{i^*}=\min_{l \in \Z_1} \ou_l$, and by choosing an $\delta_{i^*} > \delta_{j}$ (for some $j \in  \Z_1 \setminus \{i^*\}$) is inferior to the position derived by choosing a $\delta_{i^*} = \uu_s^*+\varrho+\varrho'$ (for some small enough $\varrho'$) then we have a contradiction as in the previous sub-case. 

 Otherwise, we get a $(n, \epsilon)$-external chain (with $n = |\Z_1| \ge 2$) for each $\epsilon> 0$ from   $\bdelta = (\delta_l)$ with $\delta_l = \uu_l$ for all $l \notin \Z$ and $\delta_l = \ou_l$ for all $l \in \Z $; note  $\delta_l = \ou_{i^*}$ for all $l \in \Z_1$. Starting from this $\bdelta,$ player~$i^*$ can unilaterally deviates to $\ou_{i^*} + \varrho_1$  for sufficiently small $\varrho_1 > 0$, which outperforms all other actions for player~$i^*$. That is, we have 
$$
\bdelta \to \bdelta_1 := (\delta_1, \cdots, \delta_{i^* - 1}, \ou_{i^*} + \varrho_1, \delta_{i^* + 1}, \cdots, \delta_{N}), \mbox{ where } \bdelta_1 \notin \E^o , \mbox{ and } |\{ i: \delta_{1, i} \notin \E_i^o \}| = 1.
$$
From action profile~$\bdelta_1$, any player $l \in \Z_1, \ l \neq i^* $ can deviate to $\ou_{i^*} + \varrho_2$ with $\rho_2 > \rho_1$ (slightly ahead of player~$i^*$), which again outperforms all other actions for this player~$l$. That is,
$$
\bdelta_1 \to \bdelta_2 := (\delta_1, \cdots, \delta_{l - 1}, \ou_{i^*} + \varrho_2, \delta_{l + 1}, \cdots, \delta_{N}), \mbox{ where } \bdelta_2 \notin \E^o, \mbox{ and } |\{ i: \delta_{2, i} \notin \E_i^o \}| = 2.
$$
Thus, we have $(2, \epsilon)$ chain (in similar lines one 
can have $(|\Z_1|,\epsilon)$    
external chain), for each $\epsilon >0$, which contradicts property~1. 

\medskip

\underline{To finally show that $\uu_k = \eta_{M+1}$  for each $k \le M$}

Now assume the common ${\underline u} = {\underline u}_l$ (for all $l$) is strictly bigger than $\eta_{M+1}$. Using the same construction as in \eqref{eqn_def_Z_left_end_points}-\eqref{eqn_left_end_points_contrad_arg}, we have a contradiction  again   for seller $i^*$ satisfying property 1 against $\bdelta_{-i^*}  \in \Delta_{-i^*}$: as for any $\delta \in [\underline{u}_{i^*},  {\bar u}_{i^*} ] = [\underline{u},  {\bar u}_{i^*} ]$,  and for some small enough  $\varrho' >0$ we have, 
\begin{eqnarray*}
    \U_{i^*}(\delta, \bdelta_{-i^*})  < \U_{i^*}( \eta_{M+1} + \varrho', \bdelta_{-i^*}).
\end{eqnarray*}
Observe here that the common $\uu < \teta_{M}$ and hence the above is satisfied by~\eqref{eqn_teta_def}, even if $i^* = M$. One can use arguments similar to those in the previous case for the singleton points.

\subsubsection*{Step 2: For each $i \ge k$, \ $\E_i^c \cap [\eta_{M+1}, \teta_k] =  [\eta_{M+1}, \teta_k]$ }
We have already proved that $\uu_k = \eta_{M+1}$ and that $[\eta_{M+1}, \ \eta_{M+1}+ \varrho_k] \subset \E_k^o$ with some $\varrho_k >0$ for all $k $. Now, define the following for each~$k$: 
\begin{eqnarray}
    \ug_k := \inf  (\E_k^o)^c \cap   [\eta_{M+1}, \teta_{k}] \mbox{ and hence observe }  [\eta_{M+1}, \ug_k] \subset \E_k^o.\label{Eqn_left_point_first_interval} 
\end{eqnarray}
The above is true, since $\uu_k = \eta_{M+1}$ and hence the leftmost component of any $\E_k^o$ is a closed interval in \eqref{eqn_singleton_points} and $\ug_k $ will precisely be the right endpoint of this first left component.

If  $\ug_k =   \teta_{k}$ for some $k$ then define $ \og_k = \teta_{k}$, else define:
\begin{eqnarray}\label{Eqn_right_point_first_interval}
    \og_k := \sup     \E_k^o \cap   (\ug_k, \teta_{k}]. 
\end{eqnarray} 
For all  $k$ with $\ug_k <   \teta_{k}$,   we will have $\og_k > \ug_k$ as $\E_k^o$ is a closed set; also 
 note the open interval $(\ug_k, \og_k) \subset  (\E_k^o)^c \cap   [\eta_{M+1}, \teta_{k}]$ and represents the first open-interval gap in $\E_k^o$ relative to $\E_k$. 
 
 \medskip

In the immediate next, using induction on $i$ we will show that $\min_{k\le i} \ug_k \ge \teta_{i}$; in other words we will show for every $k \leq i$ that  $[\eta_{M+1}, \teta_{i}] \in \E_k^o$. We begin with the base case, $i=M$.

\subsubsection*{Base step: to prove, $\min_k \ug_k \ge \teta_{M}$ }

For the purpose of contradiction, assume that $\min_k \ug_k < \teta_{M}$
and 
pick any $i \in \Argmin_k \ug_k$.  

\underline{If $|\Argmin_k \ug_k| < M$}, then let $j$ be a player with $\ug_j > \ug_{i}$. Now, consider
$$
\Delta = \{ \bdelta_{-i} : \delta_k \in [\ug_i -\varrho , \  \ug_i] \mbox{ for all } k \ne j, i \mbox{ and } \delta_j \in (\ug_i, \  \ug_i + \varrho)  \}  , 
$$
 for some    $\varrho > 0$.
 By choosing $\varrho $  sufficiently small (see~\eqref{Eqn_left_point_first_interval}), one can ensure  $\Delta \subset  \E_{-i}^o$  and   $\mu_{-i}(\Delta)  > 0$. Then the property~1 fails for player~$i$ against all  $\bdelta_{-i} \in \Delta$, as 
\begin{eqnarray*}
    \U_i (\ug_i + \varrho+\varrho',  \bdelta_{-i}) >  \U_i (\delta,  \bdelta_{-i})  \mbox{ for any } \delta \in \E_i^o.
\end{eqnarray*}
for some appropriately small $\varrho'>0$ (possibly depending upon $\bdelta_{-i}$); the singletons in~\eqref{eqn_singleton_points} are now placed above $\ug_i$ and one can reduce $\varrho$ further (if required) to ensure $\ug_i + \varrho+\varrho' \notin \E_i^o$. 

\medskip

 \underline{If $|\Argmin_k \ug_k| = M$}. In other words, $\ug_k = \ug_i$ for each $k $.  In this case, there exists a $\varrho > 0$ such that $(\ug_i, \ug_i+\varrho) \subset (\E_k^o)^c $ and $[\eta_{M+1}, \ug_i] \subset \E_k^o$ for each $k$  (see \eqref{Eqn_left_point_first_interval}).  Consider  $\bdelta $ with  $\delta_k = \ug_i$ for every $k$.
 One can easily construct an external $(M, \epsilon)$ chain for every $\epsilon >0$ and  smaller than a threshold, starting from this $\bdelta$. For example, consider the following chain:
 $$
 \bdelta \to \bdelta_1 := \bdelta +  (\varrho_1, 0, \cdots, 0) \to \bdelta_2 := \bdelta_1 + (0, \varrho_2, \cdots, 0) \to \cdots  \bdelta_M:= \delta_M + (\underbrace{0, \cdots, \varrho_M}_{\text{first $M$}}, \cdots, 0),
 $$
where $\varrho_i < \varrho_{i+1}$ for each $i$. This is again a contradiction. 

Hence, $\min_{k\le M} \ug_k \ge \eta_{M}$, in other words, $[\eta_{M+1}, \teta_M] \subset \E_k^o$ for every $k \leq M$ (note this implies $\E_M^o = \E_M.$)
We complete the remaining proof by induction and contradiction based arguments. 
By induction, we assume that  $\ug_k \ge \teta_{i+1}$ for all $k \leq i+1$ --- and then prove that $\ug_k \ge \teta_{i}$ for all $k \leq i$. We prove this for $i$ up to $2$, recall $\teta_1 = \teta_2$ and hence this suffices.   


\subsubsection*{Inductive step: to prove, $  \ug_k \ge \teta_{i}$ for all $k \leq i$  }

By induction  $\E_k^o = \E_k = [\eta_{M+1}, \teta_k]$ for each $k \in \{ i+1, \cdots, M\}$ and $[\eta_{M+1}, \teta_{i+1}] \subset \E_k^o$ for all $k$. For the purpose of contradiction, assume   $\min_{k\le i} \ug_k < \teta_{i}$ 
and 
pick any $l \in \Argmin_{k \le i} \ug_k$.  

\medskip

\underline{If $|\Argmin_{k \le i} \ug_k| < i$}, then let $j$ (with $j \le i$) be a player with $\ug_j > \ug_{l}$. Now consider any opponent action profile 
$$
\Delta = \{\bdelta_{-l}: \delta_k \in [\ug_l -\varrho , \  \ug_l] \mbox{ for all } k \leq l, \ k \ne j, l \mbox{ and } \delta_j \in (\ug_l, \  \ug_l + \varrho)\},
$$
for some   $\varrho > 0$.  
By choosing $\varrho > 0$  sufficiently small (see~\eqref{Eqn_left_point_first_interval}and induction-based assumption), one can ensure  $\Delta \subset  \E_{-i}^o$  and   $\mu_{-i}(\Delta)  > 0$.  Further, the property one fails for player $l$ against all such $\bdelta_{-l}$, as (recall by contradiction-based assumption,  $  \ug_l < \teta_{i} \le \teta_l$)
\begin{eqnarray*}
    \U_l (\ug_l + \varrho+\varrho',  \bdelta_{-l}) >  \U_l (\delta,  \bdelta_{-l})  \mbox{ for any } \delta \in \E_l^o,  \mbox{ and note } \ug_l + \varrho+\varrho' \notin \E_l^o,
\end{eqnarray*}
for some appropriately small $\varrho'$ (possibly depending upon $\bdelta_{-l}$).

\underline{Say $|\Argmin_{k \le i} \ug_k| = i$}. In other words, $\ug_k = \ug_l$ for each $k\le i$.  In this case, we have a $\varrho > 0$ such that $(\ug_l, \ug_l+\varrho) \subset (\E_k^o)^c $ and $[\eta_{M+1}, \ug_l] \subset \E_k^o$ for each $k \le i$  (see~\eqref{Eqn_left_point_first_interval}).  
 One can easily construct an external $(i, \epsilon)$ chain starting from $\bdelta $ with each $\delta_k = \ug_l$ for every $\epsilon$ smaller than a threshold.
 For example, consider the following chain:
 $$
 \bdelta \to \bdelta_1 := \bdelta +  (\varrho_1, 0, \cdots, 0) \to \bdelta_1 := \delta_1 + (0, \varrho_2, \cdots, 0) \to \cdots  \bdelta_i:= \delta_i + (\underbrace{0, \cdots, \varrho_i}_{\text{first $i$}}, \cdots, 0).
 $$
where $\varrho_i < \varrho_{i+1}$ for each $i$. This is again a contradiction. 

This shows that $\min_{k\le i} \ug_k \ge \teta_{i}$, i.e., $[\eta_{M+1}, \teta_{i}] \in \E_k^o$ for every $k \leq i$ and for each $i \ge 2.$
\end{proof}

\end{document}